\documentstyle[12pt,subeqnar,epsf]{article}

\begin{document}
\baselineskip=20.5pt

\def\beqra{\begin{eqnarray}} \def\eeqra{\end{eqnarray}}
\def\beqast{\begin{eqnarray*}} \def\eeqast{\end{eqnarray*}}
\def\beq{\begin{equation}}	\def\eeq{\end{equation}}
\def\be{\begin{enumerate}}   \def\ee{\end{enumerate}}

%title page
\def\fnote#1#2{\begingroup\def\thefootnote{#1}\footnote{#2}\addtocounter
{footnote}{-1}\endgroup}

\def\ut#1#2{\hfill{UTTG-{#1}-{#2}}}

\def\sppt{Research supported in part by the
Robert A. Welch Foundation and NSF Grant PHY 9009850}

\def\utgp{\it Theory Group\\ Department of Physics \\ University of Texas
\\ Austin, Texas 78712}

\def\gam{\gamma}
\def\Gam{\Gamma}
\def\la{\lambda}
\def\eps{\epsilon}
\def\La{\Lambda}
\def\si{\sigma}
\def\Si{\Sigma}
\def\al{\alpha}
\def\Tha{\Theta}
\def\tha{\theta}
\def\vphi{\varphi}
\def\del{\delta}
\def\Del{\Delta}
\def\ab{\alpha\beta}
\def\om{\omega}
\def\Om{\Omega}
\def\mn{\mu\nu}
\def\mun{^{\mu}{}_{\nu}}
\def\kap{\kappa}
\def\rsi{\rho\sigma}
\def\beal{\beta\alpha}

\def\til{\tilde}
\def\rta{\rightarrow}
\def\eqv{\equiv}
\def\nab{\nabla}
\def\pa{\partial}

\def\ul{\underline}
\def\indt{\parindent2.5em}
\def\nd{\noindent}

\def\rsi{\rho\sigma}
\def\beal{\beta\alpha}

	% calligraphic
\def\caa{{\cal A}}
\def\cb{{\cal B}}
\def\cac{{\cal C}}
\def\cd{{\cal D}}
\def\ce{{\cal E}}
\def\cf{{\cal F}}
\def\cg{{\cal G}}
\def\cah{{\cal H}}
\def\ci{{\cal I}}
\def\cj{{\cal{J}}}
\def\ck{{\cal K}}
\def\cl{{\cal L}}
\def\cm{{\cal M}}
\def\cn{{\cal N}}
\def\cO{{\cal O}}
\def\cp{{\cal P}}
\def\car{{\cal R}}
\def\cs{{\cal S}}
\def\ct{{\cal{T}}}
\def\cu{{\cal{U}}}
\def\cv{{\cal{V}}}
\def\cw{{\cal{W}}}
\def\cx{{\cal{X}}}
\def\cy{{\cal{Y}}}
\def\cz{{\cal{Z}}}

	% nots
\def\raisenot{\raise .5mm\hbox{/}}
\def\nota{\ \hbox{{$a$}\kern-.49em\hbox{/}}}
\def\notA{\hbox{{$A$}\kern-.54em\hbox{\raisenot}}}
\def\notb{\ \hbox{{$b$}\kern-.47em\hbox{/}}}
\def\notB{\ \hbox{{$B$}\kern-.60em\hbox{\raisenot}}}
\def\notc{\ \hbox{{$c$}\kern-.45em\hbox{/}}}
\def\notd{\ \hbox{{$d$}\kern-.53em\hbox{/}}}
\def\notbd{\ \hbox{{$D$}\kern-.61em\hbox{\raisenot}}} %big D
\def\note{\ \hbox{{$e$}\kern-.47em\hbox{/}}}
\def\notk{\ \hbox{{$k$}\kern-.51em\hbox{/}}}
\def\notp{\ \hbox{{$p$}\kern-.43em\hbox{/}}}
\def\notq{\ \hbox{{$q$}\kern-.47em\hbox{/}}}
\def\notW{\ \hbox{{$W$}\kern-.75em\hbox{\raisenot}}}
\def\notz{\ \hbox{{$Z$}\kern-.61em\hbox{\raisenot}}}
\def\notpa{\hbox{{$\partial$}\kern-.54em\hbox{\raisenot}}}

\def\fo{\hbox{{1}\kern-.25em\hbox{l}}}  %raised one
\def\rf#1{$^{#1}$}
\def\bx{\Box}
\def\tr{{\rm Tr}}
\def\rmtr{{\rm tr}}
\def\dgg{\dagger}

\def\lag{\langle}
\def\rag{\rangle}
\def\bmid{\big|}

\def\vlap{\overrightarrow{\La p}} %overrightarrow or left will grow over all
\def\lrta{\longrightarrow} \def\lrar{\raisebox{.8ex}{$\longrightarrow$}}
\def\rlarw{\longleftarrow\!\!\!\!\!\!\!\!\!\!\!\lrar}

\def\llra{\relbar\joinrel\longrightarrow}              %THIS IS LONG
\def\mapright#1{\smash{\mathop{\llra}\limits_{#1}}}    %ARROW ON LINE
\def\mapup#1{\smash{\mathop{\llra}\limits^{#1}}}     %CAN PUT SOMETHING OVER IT

\def\7#1#2{\mathop{\null#2}\limits^{#1}}	% puts #1 atop #2
\def\5#1#2{\mathop{\null#2}\limits_{#1}}	% puts #1 beneath #2
\def\too#1{\stackrel{#1}{\to}}
\def\tooo#1{\stackrel{#1}{\longleftarrow}}
\def\nout{{\rm in \atop out}}

\def\one{\raisebox{.5ex}{1}}
\def\BM#1{\mbox{\boldmath{$#1$}}}

\def\ltsim{\matrix{<\cr\noalign{\vskip-7pt}\sim\cr}}
\def\gtsim{\matrix{>\cr\noalign{\vskip-7pt}\sim\cr}}
\def\haf{\frac{1}{2}}

%	pictures

\def\place#1#2#3{\vbox to0pt{\kern-\parskip\kern-7pt
                             \kern-#2truein\hbox{\kern#1truein #3}
                             \vss}\nointerlineskip}

\def\illustration #1 by #2 (#3){\vbox to #2{\hrule width #1 height 0pt depth
0pt
                                       \vfill\special{illustration #3}}}

\def\scaledillustration #1 by #2 (#3 scaled #4){{\dimen0=#1 \dimen1=#2
           \divide\dimen0 by 1000 \multiply\dimen0 by #4
            \divide\dimen1 by 1000 \multiply\dimen1 by #4
            \illustration \dimen0 by \dimen1 (#3 scaled #4)}}

%-----------------------------------------
% Journal abbreviations
% -----------------------------------------------------------------------

\def\anp#1#2#3{{\it Ann. Phys. (NY)} {\bf #1} (19#2) #3 }
\def\arnps#1#2#3{{\it Ann.~Rev.~Nucl.~Part.~Sci.} {\bf #1} (19#2) #3}
\def\ijmp#1#2#3{{\it Int. J. Mod. Phys.} {\bf A#1} (19#2) #3 }
\def\jetp#1#2#3{{\it JETP Lett.} {\bf #1}  (19#2) #3 }
\def\jpb#1#2#3{{\it J. Phys.} {\bf B#1}  (19#2) #3}
\def\mpla#1#2#3{{\it Mod.~Phys.~Lett.} {\bf A#1}  (19#2) #3}
\def\nci#1#2#3{{\it Nuovo Cimento} {\bf #1} (19#2) #3}
\def\npb#1#2#3{{\it Nucl. Phys.} {\bf B#1} (19#2) #3 }
\def\plb#1#2#3{{\it Phys. Lett.} {\bf B#1} (19#2) #3 }
\def\prc#1#2#3{{\it Phys. Rev.} {\bf C#1} (19#2) #3 }
\def\prd#1#2#3{{\it Phys. Rev.} {\bf D#1} (19#2) #3 }
\def\pr#1#2#3{{\it Phys. Rev.} {\bf #1} (19#2) #3 }
\def\prep#1#2#3{{\it Phys. Rep.} {\bf C#1} (19#2) #3 }
\def\prl#1#2#3{{\it Phys. Rev. Lett.} {\bf #1}(19#2) #3 }
\def\rmp#1#2#3{{\it Rev. Mod. Phys.} {\bf #1} (19#2) #3 }
\def\zpc#1#2#3{{\it Zeit. Phys.} {\bf #1} (19#2) #3 }

\def\ON{{\cal O}(N)}
\def\UN{{\cal U}(N)}
\def\bdPh{\mbox{\boldmath{$\dot{\!\Phi}$}}}
\def\bPh{\mbox{\boldmath{$\Phi$}}}
\def\bPhs{\bPh^2}
\def\sef{S_{eff}[\sigma]}
\def\sigx{\sigma(x)}
\def\bph{\mbox{\boldmath{$\phi$}}}
\def\bphs{\bph^2}
\def\ex{\BM{x}}
\def\exs{\ex^2}
\def\xdot{\dot{\!\ex}}
\def\y{\BM{y}}
\def\ys{\y^2}
\def\ydot{\dot{\!\y}}
\def\pat{\pa_t}
\def\pax{\pa_x}

\renewcommand{\thesection}{\Roman{section}}
\ut{03}{94}

\hfill{hep-th/9404143}

\hfill{March 1994}

\vspace*{.3in}
\begin{center}
  \large{\bf Non Trivial Saddle Points and Band Structure of Bound \\[6pt]
States of the Two Dimensional \mbox{\boldmath${\cal O}(N)$} Vector Model}
\normalsize

\vspace{36pt}
Joshua Feinberg\fnote{*}{ Supported by a post doctoral Rothchild Fellowship
and in part by the Robert A. Welch Foundation and NSF Grant PHY 9009850.}

\vspace{12pt}
{\it Theory Group, Department of Physics\\
The University of Texas at Austin, RLM5.208, Austin, Texas 78712\\

\vspace{4pt}
e-mail joshua@utaphy.ph.utexas.edu}

\vspace{.6cm}

\end{center}

\begin{minipage}{5.3in}

{\abstract~~~~~We discuss $\ON$ invariant scalar field theories in $0+1$ and
$1+1$ space-time dimensions.  Combining ordinary ``Large $N$" saddle point
techniques and simple properties of the diagonal resolvent of  one
dimensional Schr\"odinger operators we find {\it exact} non-trivial (space
dependent) solutions to the saddle point equations of these models in addition
to the saddle point describing the ground state of the theory.  We interpret
these novel saddle points as collective $\ON$ singlet excitations of the field
theory, each embracing a  host of finer quantum states arranged in $\ON$
multiplets, in an analogous manner to the band structure of molecular spectra.
We comment on the relation of our results to the classical work of Dashen,
Hasslacher and Neveu and to a previous analysis of bound states in the $\ON$
model by Abbott.}

\end{minipage}

\vfill
\pagebreak

\setcounter{page}{1}

\section{Introduction}
\renewcommand{\theequation}{1.\arabic{equation}}

\indent\indent Field theories involving a very large number $N$ of components
have turned out to be an extremely useful tool in addressing many
non-perturbative aspects of quantum field theory and statistical mechanics
\cite{col,berwad,brez}.  In such theories the action is proportional to $N$
(or in some cases, to $N^2$) which plays the role of $1/\hbar$.  Therefore, as
$N\rta\infty$, the path integral of such field theories is dominated by the
saddle points of the action, which are protected against being washed out by
quantum fluctuations in this limit.

An important subclass of such field theories are the $\ON$ vector models whose
dynamical variables are $N$ real scalar fields arranged into an $\ON$ vector
$\BM\Phi$ with $\ON$ invariant self interactions of the form $(\BM{\Phi}^2)^n$
\footnote{We concentrate on the case of quartic interactions $n=2$} which have
been studied intensively in the past \cite{larn,aksh,sh,mb}.  More recently,
$\ON$ vector models have attracted interest \cite{zj,dvc,mdvc,sch,mm} in
relation to the so called ``double scaling limit''
\cite{brez}.  The cut-off dependent effective field theory has also been
discussed very recently
\cite{schnu}.  The main object of these studies were the true vacuum state of
the $\ON$ model, and the small quantum fluctuations around it.  In the case of
quartic interactions, of interest to us here, the theory is best analyzed by
introducing into the lagrangian an auxiliary field $\sigma$ that can be
integrated out by its equation of motion $\sigma${}$ = \frac{g}{2N}\,\bPhs$
(where
$g^2/N$ is the quartic coupling in the lagrangian).\footnote{For higher
interactions we need an additional lagrange multiplier field.} In this way the
original fields $\bPh$ may be integrated out exactly, yielding an effective
non-local action $\sef$ in terms of $\sigma$ alone.

The vacuum and the low lying states above it are described by the small
fluctuations of the $\si$ field around a certain local minimum of
$\sef$\footnote{This minimum may not be the {\it absolute} minimum of $\sef$ in
certain circumstances \cite{mb}, \cite{col} but the decay rate of the
metastable vacuum is suppressed by a very small tunnelling factor of the order
of $e^{-NV}$, where $V$ is the volume of space-time.}.
In this minimum the $\si$ field configuration is homogeneous (i.e. space-time
independent) $\si=\si_c$, where the constant $\si_c$ (which is the true mass of
the $\bPh$  ``mesons" around that vacuum) is obtained from the gap equation
$\frac{\delta\sef}{\delta\si}\big|_{\si=\si_c}=0$.
Homogeneity of the vacuum $\si$ configuration is expected due to Poincar\'e
invariance of the latter.

$\sef$ is a rather complicated functional of $\si$ and in addition to
$\si=\si_c$ it possesses yet many other extremal points at which the solution
to
the extremum condition $\frac{\del\sef}{\del\si}=0$ are space-time dependent
$\si$ configurations $\sigx$.
These extrema cannot describe the ground state (and indeed, neither a
metastable ground state) and are thus {\it saddle points} of $\sef$ rather than
(local) minima.  Indeed, it can be shown by direct calculation\footnote{ This
will not be carried out here.}  that the second variation around them has an
infinite number of  negative eigenvalues.  They correspond to collective
``heavy'' $\ON$ singlet excitations of the $\bPh$ field, where its length
squared oscillates in isotopic space as a function of space and time.

In this work we find explicit {\it exact} formulae for these $\sigx$
configurations in $0+1$ and in $1+1$ space time dimensions.  In the two
dimensional case we discuss only time independent $\si$ configurations.  In
either dimensionalities they have the general form $\sigx=\psi(x-x_0)$, where
$\psi$ is an elliptic function and $x_0$ is an arbitrary parameter responsible
for translational invariance.  In $0+1$ dimensions  if the amplitude of
$\psi(t)$ along the real axis is finite it corresponds to a finite $\sef$
(after regularizing the contribution of the fluctuations of the $\ON$ vector
field into a finite expression), while if the amplitude is infinite (i.e. the
pole of $\psi$ lies on the real axis) it is a potential source of  an infinite
$\sef$, which can suppress this mode enormously relative to the homogeneous
$\si=\si_c$ configuration.  Moreover, such infinite amplitude
$\si (t)$ configurations, imply very large  fields, a fact that might
throw us away from the validity domain of the large $N$ approximation.  We will
always keep this in mind when addressing such configurations, but in $0+1$
dimensions we argue that they might be relevant in certain circumstances.
In the two dimensional field theory case most $\sigx=\psi(x-x_0)$
configurations give rise to infinite $\sef$ values (whose divergences are
inherent and cannot be removed by regularization) and are therefore highly
suppressed.  Infinite amplitude $\sigx$ configurations fall obviously into
this category, but so do finite amplitude $\sigx$ configurations.  Some of
the latter yield (regularized) finite $\sef$ values only in the limit in which
their real period (as elliptic functions) becomes infinite.  Similar $\sigx$
configurations were found in the two dimensional
$\ON$ model in \cite{abbott} by using inverse scattering techniques.
Nevertheless, they turn out to be quite distinct from those discussed here.
Our finite amplitude periodic
$\sigx$ configurations might become important if we put the field theory in a
spatial ``box'' of a finite length
$L$, but as $L\rightarrow \infty$ they will be hightly suppressed due to
their diverging $\sef$.

These $\sigx$ saddle point configurations are hybrid objects, in the sense that
they are static solutions of the {\it classical} equations of motion
$\frac{\del\sef}{\del\si}=0$ of the action $\sef$ that embodies {\it all}
quantal effects of the $\bPh$ field.  Therefore, due to all their properties
enumerated above, the finite $\sef\;~\sigx$ configurations are reminiscent, in
some sense, of soliton solutions in other field theories, but this analogy is
only a remote one, since the $\sigx$'s carry no topological charge of any sort
and they are {\it not} local minima of $\sef$.  They are more similar to the so
called ``non-topological solitons'' \cite{tdl}, \cite{will}.  In this paper we
ignore the gaussian fluctuations of the $\si$ field around the saddle points
$\sigx$, but explain how to find the spectrum of ``mesons" (i.e. $\bPh$
quanta) coupled to the $\sigx$ configuration.
 It is clear  that each $\sigx$ configuration gives rise to a tower of
``meson"
bound states as well as to ``meson" scattering states, arranged in highly
reducible  $\ON$ multiplets. The infinite negative eigenvlues of the second
variation of $\sef$ around these configurations correspond to the infinite
number of decay modes of such excited states into all lower ones. This form of
the spectrum of quantum states is reminiscent of the vibrational-rotational
band structure of molecular spectra
\cite{col}.  Since we ignore fluctuations of the $\si$ field around the saddle
point configuration $\sigx$, the ``molecular analog'' to this would be to have
the molecule only at one of its vibrational ground states and consider the
rotational excitations around it.  As a matter of fact, in the $0+1$
dimensional case, which is nothing but the quantum mechanics of a single
unharmonic oscillator in $N$ euclidean dimensions this analogy is actually the
correct physical picture.  In this case the $\si(t)$ configurations
 are actual vibrations of the $N$ dimensional
vector,  changing its length.  As the vector stretches or contracts, it
might rotate  as well, changing its orientation in $N$ dimensional space,
which is energetically much cheaper.  The latter gives rise to the rich $\ON$
structure of the spectrum.

Our approach to the specific problem of bound state spectrum in the $1+1$
dimensional $\ON$ model has been inspired by the seminal papers of Dashen,
Hasslacher and Neveu (hereafter abbreviated as DHN) \cite{dhn1}, \cite{dhn2},
\cite{dhn3},  and especially, by their analysis of the bound state spectrum in
the Gross-Neveu model \cite{gn}.  In \cite{dhn3} DHN present a detailed
construction of a sector of the bound state spectrum of the Gross-Neveu model
whose corresponding $\sigx$ configurations\footnote{Here $\sigx$ stands for the
$\bar\psi\psi$ bilinear fermionic condensate. We refer here only to their
static $\sigx$ configurations.} differ only slightly from the homogeneous
vacuum configuration $\si=\si_c$ of the Gross-Neveu model, at least as far as
states with a  principal quantum number well below $N$ are concerned.  For
such states, the corresponding $\sigx$ configuration looks like a pair of
interacting (very close) kink and anti-kink, where the kink amplitude is of
the order $1/N$.  This might be thought of as the back-reaction of the
fermions in the bound state on the vacuum
$\si=\si_c$ polarizing it into a $\sigx$ configuration.  Despite the smallness
of $\sigx-\si_c$, it has a dramatic consequence: the production of its
corresponding bound state.  Indeed, DHN find the explicit form of $\sigx$ by
performing an inverse scattering analysis of the one dimensional Schr\"odinger
operator $\partial_x^2+\sigx^2-\si_c^2+\si'(x)$ obtained straightforwardly from
the Dirac operator $i\notpa-\si$. The potential term in this Schr\"odinger
operator vanishes identically and cannot bind for $\si =\si_c$, but for the
$\sigx$ configuration of DHN it becomes an attractive reflectionless
potential (of depth of the order $1/N^2$).  Thus, despite the small magnitude
of
the ``back-reaction'' of the fermions on the vacuum, it is enough in order to
create a very shallow dip in the one dimensional Schr\"odinger potential,
causing it to produce one bound state.  This is the reason why DHN attribute
the appearance of fermionic bound states in the Gross-Neveu model to its
intrinsic infra-red instabilities (the distortion of $\si_c$ into $\sigx$)
which
result from the asymptotic freedom of the model. Indeed, the same infra-red
instabilities generate non-perturbatively the fermion mass $g\si_c$ in this
model in the first place.

The $1+1~~\ON$ vector model with quartic interactions being trivially
``asymptotically free" due to its super-renomalizability, exhibits for negative
quartic coupling a sector of bound states (whose $\sigx$ configuration was
mentioned above) that is  very similar to the bound states of the Gross-Neveu
model. It is probably created by the same mechanism of vacuum infra-red
instabilities.  Their energy spectrum and corresponding reflectionless kink
like $\sigx$  configurations were found in
\cite{abbott} by following the DHN	prescription step by step.

In addition to the description above, DHN mention briefly yet another sector of
bound states \cite{dhn3} in the Gross-Neveu model, namely, those built around
the Callan-Coleman-Gross-Zee kink.  The latter is a local mimimum $\sigx$
configuration very different from the homogeneous vacuum one.  It connects the
two true degenerate vacua  of the Gross-Neveu model.
 Here all fermions are bound at zero binding energy in
the fermionic zero mode of the kink (which is the only bound state supported by
the kink),  with
{\it no back-reaction at all} on the kink independently of the number of
fermions trapped in that state.  The
$\sigx$ configurations we find  in the $\ON$ model are precisely of this type.
However, the latter cannot be associated with degenerate vacua, since the
$\ON$ model lack such a structure.
Due to this fact and since we do not use inverse scattering techniques, our
results are complementary to those of \cite{dhn3,abbott}.  Our method
could be used also to find analogues to the Callan-Coleman-Gross-Zee kink in
other two dimensional theories, where inverse scattering methods become
practically useless
\cite{sss}.
As far as we know this is the first use
made of the method described below.  The idea we use is very simple.  A generic
saddle point condition for the $\ON$ model (in any dimension) $
\frac{\del\sef}{\del\si}=0$ relates $\si$ to the diagonal resolvent of a
Klein-Gordon operator (for scalar fields)  with $\si$ as its potential term.
In the static case the Klein-Gordon operator reduces into a Schr\"odinger
operator.  In
$1+1$ dimensions the resulting diagonal resolvent of the one dimensional
Schr\"odinger operator is known to obey a certain differential equation, known
as the Gelfand-Dikii equation \cite{gd}.  Since the resolvent is essentially
$\sigx$ due to the saddle point condition, the latter must obey an ordinary
differential equation induced from the Gelfand-Dikii equation.

The differential equation imposed on $\sigx$ is of second order.  Solving it
for $\sigx$ we encounter two integration constants.  One constant is trivial
and
insures translational invariance of the solution for $\sigx$ while the other
acts as a  parameter that distorts the shape of $\sigx$ and parametrises its
period along the $x$ axis.  When $\sigx$ has a finite amplitude we can always
use this parameter to drive  $\sigx$ into an infinite period behaviour.  In
this
limit $\sigx$ reproduces the reflectionless  configuration analogous to the
Callan-Coleman-Gross-Zee kink in the case of negative quartic coupling.

At this point it is worth mentioning that some bound states in $\ON$ models in
various dimensions were found around the homogeneous vacuum $\si=\si_c$
configuration in the singlet $\bPh\cdot\bPh$ channel of ``meson-meson"
 scattering,
by simply analyzing the poles of the $\si-\si$ propagator \cite{aksh},
\cite{sh},
\cite{mb}.\footnote{See also the third reference in \cite{larn}.} Specific
conditions for the existence of such bound states and resonances in two
space-time dimensions to four may be found in \cite{aksh,sh}.   In certain
theories  in three space-time dimensions such bound states may become massless
(to leading order in $1/N$) and are identified as goldstone poles (a dilaton
\cite{bmb} or a dilaton and a dilatino
\cite{bhm}) associated with spontaneous scale invariance breakdown.  Occurance
of such massless bound states is responsible for the double scaling limit in
vector models \cite{mm} and the consequences of obstructions for
this to happen in two space-time dimensions were addressed in \cite{mdvc}.

The paper is organized as follows:  In section (II) we analyze the $0+1$
dimensional model.  This is done mainly to introduce notations and gain
confidence in our method.  In particular, we show that our method for obtaining
the time dependent saddle point $\si$ configuration is completely equivalent
to a JWKB analysis of the radial Schr\"odinger equation of the unharmonic
oscillator where $\hbar=1/N$ establishing the validity of our method.  We
analyze both cases of real and imaginary time qualitatively, without giving
explicit formulae for the $\si(t)$ configurations.

In section (III) we turn to the $1+1$ dimensional case.  Here as well we
analyze both Minkowsky and Euclidean signatures of space-time.  Our static
$\sigx$ configurations are obtained in a certain adiabatic approximation to
the saddle point equation.  Non-zero frequency modes
are strongly suppressed and  the only meaningful result of the saddle
point equation concerns the zero frequency mode.  In this way we effectively
reduce the theory from $1+1$ dimensions to $0+1$ dimensions.  We give an
explicit expression for some of  the $\sigx$ configuration  that is important
in the infinite volume case and construct the spectrum of ``mesonic" bound
states in its background.

We draw our conclusions in section (IV).  A simple proof of the Gelfand-Dikii
equation, emphasizing its elementarity in the theory of one dimensional
Schr\"odinger operators is given in an appendix.

\pagebreak
\renewcommand{\theequation}{2.\arabic{equation}}
\section{The Unharmonic Oscillator}
\setcounter{equation}{0}

\indent\indent We begin our investigation of the bound state spectrum in the
two dimensional $\ON$ vector model by considering the simpler case of the
quantum mechanics of a single $\ON$ invariant unharmonic oscillator.  Clearly,
the most straightforward way of analyzing the quantum state spectrum of this
system when
$N\rightarrow \infty$ is to apply the large $N$ approximation directly to the
hamiltonian of this oscillator.  This procedure is entirely equivalent to the
JWKB approximation to the corresponding radial Schr\"odinger equation, where
$\hbar=\frac{1}{N}$ \cite{witt}.  Comparing our method of calculating the
spectrum, which is an analysis of the action in the large $N$ limit, to the
JWKB approximated Schr\"odinger equation reveals their equivalence, which is
not surprising, since the large $N$ approximation to the path integral is
equivalent to the semiclassical approximation with the role of $\hbar$ played
by $1/N$.

Therefore, our purpose in this section is to  better understand  the
way our method works in this simpler case and gain confidence in it before
generalizing to the field theoretic case.  Such an understanding  is
achieved precisely by demonstrating  its equivalence to the JWKB approximated
Schr\"odinger equation. This reasoning is completely analogous to the one made
in \cite{pol,polbook,col}, in order to understand the role of instantons as
tunnelling configurations in the path integral.  Indeed, tunnelling effects in
quantum mechanics are calculated straightforwardly by using JWKB approximated
Schr\"odinger equations rather  than path integral instanton methods, but it
is the instanton calculus that is generalizable to quantum field theory and
not the JWKB approximated  Schr\"odinger equation.
Since our main interest in this paper is the $1+1$ dimensional field
theoretic case (discussed in the next section) we will demonstrate the
consistency of our method in the $0+1$ dimensional case without giving the
resulting $\si(t)$ configurations explicitly in terms of elliptic functions of
$t$.  Evaluation of these functions may be done straightforwardly from our
discussion.

The $\ON$ invariant unharmonic oscillator is described by the
action\footnote{We assume throughout this work that $m^2\geq 0$.  A sign flip
of $m^2$ is equivalent to a simultaneous sign flip of the parameters $\al$ and
$\beta$.}
\beq
S=\int\limits^T_0\, dt \left[\haf\,\xdot^2-\al\left(\haf\,m^2 \ex^2 +
\frac{1}{4N}\,\beta g^2(\ex^2)^2\right)\right] \label{2una}
\eeq
where $\ex$ is an $N$ dimensional vector, $g^2$ is the quartic coupling and $m$
is the pure harmonic frequency (as $g=0$) of the oscillator.  Here $\al$ and
$\beta$ are discrete parameters whose values are either $+1$ or $-1$.  $\al=1$
corresponds to the real time case, while $\al=-1$ corresponds to imaginary
(euclidean) time.  Likewise, $\beta=1$ corresponds to the well defined stable
case where the quartic potential goes to $+\infty$ as $|x|\rightarrow \infty$,
while $\beta=-1$ corresponds to the upside down potential.  We have introduced
these parameters for efficiency reasons, allowing us to discuss all different
possibilities compactly.  In the large $N$ approximation
$g^2$ is finite, hence the quartic interactions in Eq. (\ref{2una}) are in the
weak coupling regime\footnote{Taking the scale dimension of $m$ to be
canonically 1 the other dimensions are $[t]=-1,\;[\ex]=-\haf,\;[g^2]=3$.  The
dimensionless ratio $\frac{g^2/N}{m^3}$ which is much smaller than 1 for finite
$g^2,m^2$ clearly implies weak coupling.}.

Rescaling
\beq
\ex^2=N\ys \label{2dos}
\eeq
Eq. (\ref{2una}) may be rewritten as
\beq
S=N \int\limits^T_0 dt\left[\haf\;\ydot^2-\al\;\left(\frac{m^2}{2}\;\ys+
\frac{\beta g^2}{4}\,(\ys)^2\right)\right]\,.  \label{2tres}
\eeq

Introducing an auxiliary variable $\sigma$, the action is finally recast into
\beq
S=N\int\limits^T_0 dt \left[\haf\;\ydot^2-\frac{\al m^2}{2}\, \ys +
\ab(\sigma^2-g\sigma\ys)\right]  \label{2qrto}
\eeq
which is the starting point of our calculations.  Note that $\sigma $ has no
kinetic energy and thus may be eliminated from Eq. (\ref{2qrto}) via its
equation of motion
\beq
\sigma =\haf \,g\,\ys\;.  \label{2cin}
\eeq

Throughout this paper we adopt the convention $g>0$ and therefore $\sigma $
is a non-negative  variable.

Since in principle  we are interested in calculating the spectrum of bound
state
of the oscillator we consider the partition function $\tr\,
e^{\frac{1}{\gam}\, H\,T} =\int d^N \y_0\, W_P(T)$ where $W_P(T)$ is
the transition amplitude of the oscillator to start at
$\y_0$ and return there after time-lapse $T$
\beq
\lag\y_0;t=T|\y_0;\, t=0\rag\eqv W_P(T) = \int\limits_{pbc}\cd\y\cd\sigma \,
e^{\gamma S}\;. \label{2siz}
\eeq
Here the subscript ``$pbc$" denotes integration over paths obeying the obvious
periodic boundary conditions implied by Eq. (\ref{2siz}) and $\gamma=i$ for the
real time case, while $\gamma=-1$ for imaginary time.  Clearly, the periodic
boundary conditions on $\si$ are dictated by Eq. (\ref{2cin}).

Eq. (\ref{2qrto}) is quadratic in $\y$ which may be therefore integrated
completely out of the action leading to
\beq
W_P(T)=\int\limits_{pbc}\cd\sigma\,\det\nolimits^{-N/2}_{pbc} \left[-\pat^2-\al
m^2-2\ab g\sigma\right] e^{N\ab\gamma\int^T_0\sigma^2 dt}\label{2set}
\eeq
where the operator whose determinant is taken in Eq. (\ref{2set}) is defined on
the interval $[0,T]$ with periodic boundary conditions.  Exponentiating the
determinant in Eq. (\ref{2set}) we may express $W_P(T)$ as
\beq
W_P(T) = \int\limits_{pbc}\cd\sigma\, e^{N\gamma\sef} \label{2och}
\eeq
where
\beq
\sef = \ab\int\limits_0^T \sigma^2 -\frac{1}{2\gamma}\, \mbox{Tr}_{pbc}
\ln\left[-\pat^2- \al m^2-2\ab g\sigma\right]   \label{2nev}
\eeq
is the exact effective non-local action in terms of the $\sigma$ field.

Due to the explicit factor of $N$ in the exponent of Eq. (\ref{2och}), $W_P(T)$
is dominated by the extremal points of $\sef$, and to leading order in
$1/N$ \footnote{Thus ignoring quadratic fluctuations of $\sigma$ around these
extrema.}  it is given by the {\it coherent} sum over all such extremal
configurations $\sigma(t)$ of the integrand in Eq. (\ref{2och}) evaluated at
these configurations \cite{ls}.{} \footnote{In case of finite dimensional
integrals, not all saddle points necessarily contribute.  Only those that are
located along the steepest contour contribute
\cite{benor}.  In the functional integral (Eq. (\ref{2och})) the situation is
not so clear.  However all $t$ dependent extrema $\si(t)$ we find have simple
classical meaning, and therefore must be all important.}

These extremum configurations are solutions of the functional  condition
\beq
\delta\sef/\delta \sigma=0~~;~~\sigma(0) =\sigma(T)   \label{2dez}
\eeq
which is equivalent to
\beq
2\sigma(t)=-\frac{g}{\gamma}\lag t|\frac{1}{-\pat^2-\al m^2-2\ab
g\si}\;|t\rag;\;\;\si(0)=\si(T)\;.   \label{2onte}
\eeq

Eq. (\ref{2onte}) is a rather complicated functional equation for generic $\si$
configurations.  However, it simplifies enormously for {\it constant}
$\si=\si_c$ configurations which obey the periodicity boundary condition
automatically for any $T$.  In this case the diagonal matrix element is given
by a simple Fourier integral (for $T\rta\infty$), and Eq. (\ref{2onte})
descends
into the cubic equation
\beq
32\beta g\si_c^3+16 m^2\si_c^2-g^2=0\,. \label{2doce}
\eeq
Eq. (\ref{2doce}) has been used  in \cite{ben,zj,dvc} \footnote{See also the
third reference in \cite{larn}.} to determine the optimal frequency squared
$\om_c^2=m^2+2\beta g\si_c$ of the harmonic approximation to Eq. (\ref{2una})
for
low lying excitations around the vacuum (or metastable vacuum, for $\beta=-1$)
of  the unharmonic oscillator.

Following Eq. (\ref{2cin}) one should obviously consider only {\it positive
real}  roots of Eq. (\ref{2doce}).  Analyzing the left hand side of Eq.
(\ref{2doce}) it is clear that for $\beta=+1$ there is only one such root,
while
for $\beta=-1$ there are two, but only the smaller one corresponds to the
metastable vacuum.\footnote{In this case
$\om_c^2=m^2-2g\sigma_c$ and $\om_c^2<0$ for the larger root.}  These
assertions will be demonstrated more clearly when we will examine the JWKB
approximation to the radial Schr\"odinger equation of the unharmonic
oscillator in the last part of this section.

We now turn to generic time dependent solutions of Eq. (\ref{2onte}).  This
equation relates the diagonal resolvent $R(t)=\lag t|(h-\al m^2)^{-1}|t\rag$ of
the Schr\"odinger operator
\beq
h=-\pat^2-2\ab g\sigma   \label{2dtre}
\eeq
whose potential term is $U(t)=-2\ab g\sigma$, to $\sigma(t)$.  This diagonal
resolvent obeys the Gelfand-Dikii equation (Eq. (A.12)) discussed in
the appendix.  Thus, eqs. (\ref{2onte}) and (A.12) imply
\beq
2\si\si{''}-(\si')^2+4\al(m^2+2\beta
g\si)\si^2=\frac{g^2}{4\al}\,;\;\;\si(0)=\si(T)  \label{2dqrt}
\eeq
where $\si'=d\si/dt$ and we have used the relation $\al=-\gam^2$.

Eq. (\ref{2dqrt}) is one of the main results of this paper.  It implies that
the {\it functional} equation Eq. (\ref{2onte}) is equivalent to an {\it
ordinary  differential} equation that we readily solve below, obtaining the
required extremal $\si(t)$ configuration.

Note that Eq. (\ref{2dqrt}) admits a time independent solution $\si=\si_c$ for
which it reproduces Eq. (\ref{2doce}) exactly.  Regarding time dependent
solutions, first integration of Eq. (\ref{2dqrt}) may be done straightforwardly
by substituting
\beq
\si'(t)=f(\si)    \label{2dcin}
\eeq
which leads to
\beq
f^2(\si)\eqv\left(\frac{d\si}{dt}\right)^2
=\al\left[\frac{g^2}{4}\,\left(\frac{\si}{\si_0}-1\right) -4
\left(m^2\si^2+\beta
g\si^3\right)\right];\; \si(0)=\si(T)   \label{2dsiz}
\eeq
where $\si_0$ is an integration constant (and we have used $\al^2=1$).  Eq.
(\ref{2dsiz}) has the form of the equation of motion of a point particle in one
dimension along the  ray\footnote{Here we assume that the constraint $\si\geq
0$ stemming from the classical equaiton of motion of $\si$ (Eq. (\ref{2cin})
remains in tact quantum mechanically.  We will see that this is indeed the
case.
Since Eq. (\ref{2cin}) could be enforced as an exact {\it functional}
constraint
by introducing a lagrange multiplier field, this conclusion should not be of
any surprise.}
 $\si \geq 0$ in the potential $V(\si)$ given by
\beq
-V(\si)=\frac{g^2}{4}\left(\frac{\si}{\si_0}-1\right) -4\left(m^2\si^2+\beta
g\si^3\right);~\si\geq 0   \label{2dset}
\eeq
(we have shifted the constant piece of $V$ such that Eq. (\ref{2dsiz})
corresponds to a motion at energy zero).  Note that $\al$ has been factored out
in front of
$-V$ in Eq. (\ref{2dsiz}), as it should, since $\al=+1$ corresponds to motion
in real time while $\al=-1$ corresponds to motion in imaginary time.  The
classical turning points are the {\it non-negative} roots of the cubic equation
\beq
V(\si)=0\,.  \label{2doch}
\eeq

For real time motions, the classical $\si$ paths will be confined to regions
where $-V\geq 0$, while for imaginary time motions, to regions where $-V\leq
0$.
In addition to the turning points given by Eq. (\ref{2doch}), there is a
turning point at
$\si=0$ due to the constraint $\si\geq 0$.\footnote{$\si$ trajectories that
reach the endpoint $\si=0$ always occur in imaginary time.  They are
undesirable since they have no quantal counterpart.  We comment on this point
in the last part of this section.}

Denoting the roots of Eq. (\ref{2doch}) by $\si_1,\si_2,\si_3$ \footnote{We
adopt the convention that when all three roots are real they are ordered as
$\si_1\leq\si_2\leq\si_3$.} the following formulae are useful for listing the
various solutions of Eq. (\ref{2dsiz}).
\begin{subeqnarray}
-V(0) &=& - \frac{g^2}{4}< 0  \label{2dnev}\\
-V(\pm\infty) &=& \mp \beta\cdot\infty   \label{2dnev2}\\
\si_1\si_2\si_3 &=& - \frac{\beta g}{16}\;. \label{2dnev3}
\end{subeqnarray}
  Assuming $m$ and
$g$ are given, $\si_0$ is the only arbitrary parameter, arising as an
integration constant of Eq. (\ref{2dsiz}).  We now show that equation
(\ref{2dsiz}) is equivalent to the (one dimensional) radial equation of motion
of the oscillator in a {\it quantum effective} potential obtained as
$N\rightarrow
\infty$, whose energy parameter $\ce $ is inversely proportional to $\si_0$.

This observation is conceptually an important one, assuring the validity of our
calculation of the time dependent extremal $\si$ configurations of $\sef$,
identifying them as corresponding to collective radial excitations of the
oscillator, as we have discussed at the beginning of this section.  Moreover,
holding $m$ and $g$ fixed, $\si_1,\si_2$ and $\si_3$ are functions of $\si_0$,
and in the most generic case two of them are complex (conjugate).  Therefore,
the best way to analyze the various turning point configurations of the
solutions to Eq. (\ref{2dsiz}) is to consider the (large $N$) effective radial
potential of the oscillator giving rise to these radial motions.  Varying the
energy parameter of the latter will show us the way the turning points of Eq.
(\ref{2dsiz}) change as functions of $\si_0$. Recall that $W_P(T)$ defined in
Eq. (\ref{2siz}) is given as a sum over all saddle points $\si(t)$ found from
Eq. (\ref{2onte}).  These are essentially parametrized by $\si_0$.  Thus a
crude
estimate of $W_P(T)$ would be to insert these saddle point configurations into
the integrand of Eq. (\ref{2och}) and sum over all allowed values of $\si_0$.
Since
$\si_0$ turns out to be inversely proportional to the energy parameter $\ce$
of the associated radial Schr\"odinger equation this is just the natural
thing to do, since by definition $W_P(T)$ is a Laplace transform of the
transition amplitude
$W_P(\ce)$ in the energy plane.  In what follows, howerver, we will not
calculate $W_P(T)$ but rather concentrate on a specific saddle point $\si(t)$
with its particular $\si_0$ parameter.

Before turning to the hamiltonian formulation we depict qualitatively in
figures (1) and (2) the turning point structure of Eq. (\ref{2dsiz}) in the
case where
$\si_1,\si_2$ and $\si_3$ are all {\it real}.  Figure (1) corresponds to the
case $\beta=+1$ while Figure (2) describes the various possibilities for
$\beta=-1$.
When $\beta=+1$, Eqs. (\ref{2dnev}) imply $-V(0)<0, ~-V(\mp\,\infty)=\pm\infty$
and
$\si_1\si_2\si_3<0$.  Therefore, if $\si_1,\si_2$ and $\si_3$ are all real,
either $\si_1<\si_2<\si_3<0$ or $\si_1<0<\si_2<\si_3$.  These two possibilities
are shown in figures (1a) and (1b), respectively.
For $\beta=-1$ we have $-V(0)<0,\;-V(\pm\infty)=\pm\infty$ and
$\si_1\si_2\si_3>0$.  Thus, if $\si_1,\si_2$ and $\si_3$ are all real either
$\si_1<\si_2<0<\si_3$ or $0<\si_1<\si_2<\si_3$.  These two possibilities appear
in figures (2a) and (2b), respectively.

\pagebreak

\begin{center}
\epsfxsize=0in\hspace*{0in}
\epsffile{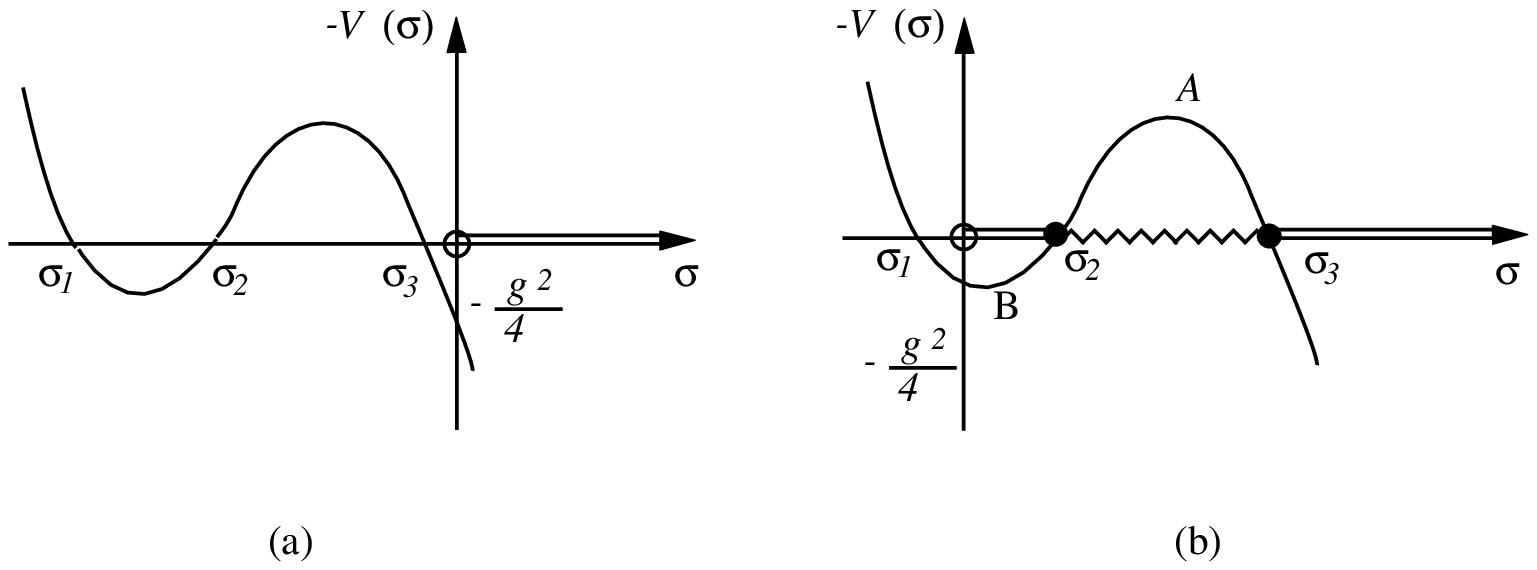}
\end{center}

\vspace{12pt}\nd
FIGURE 1: ~ {The $\beta=1$ case of Eq. (\ref{2dsiz}).  Either
$\si_1<\si_2<\si_3<0$ (a) or $\si_1<0<\si_2<\si_3$ (b).  Bold faced dots
correspond to turning points of the $\si(t)$ motions.  The empty circle denotes
the reflecting infinite potential wall at $\si=0$.  The regions along the $\si$
axis accessible to real time motions are marked by the zig-zag line and those
accessible to motions in imaginary time-by double lines.}

\vspace{16pt}
\begin{center}
\epsfxsize=0in\hspace*{0in}
\epsffile{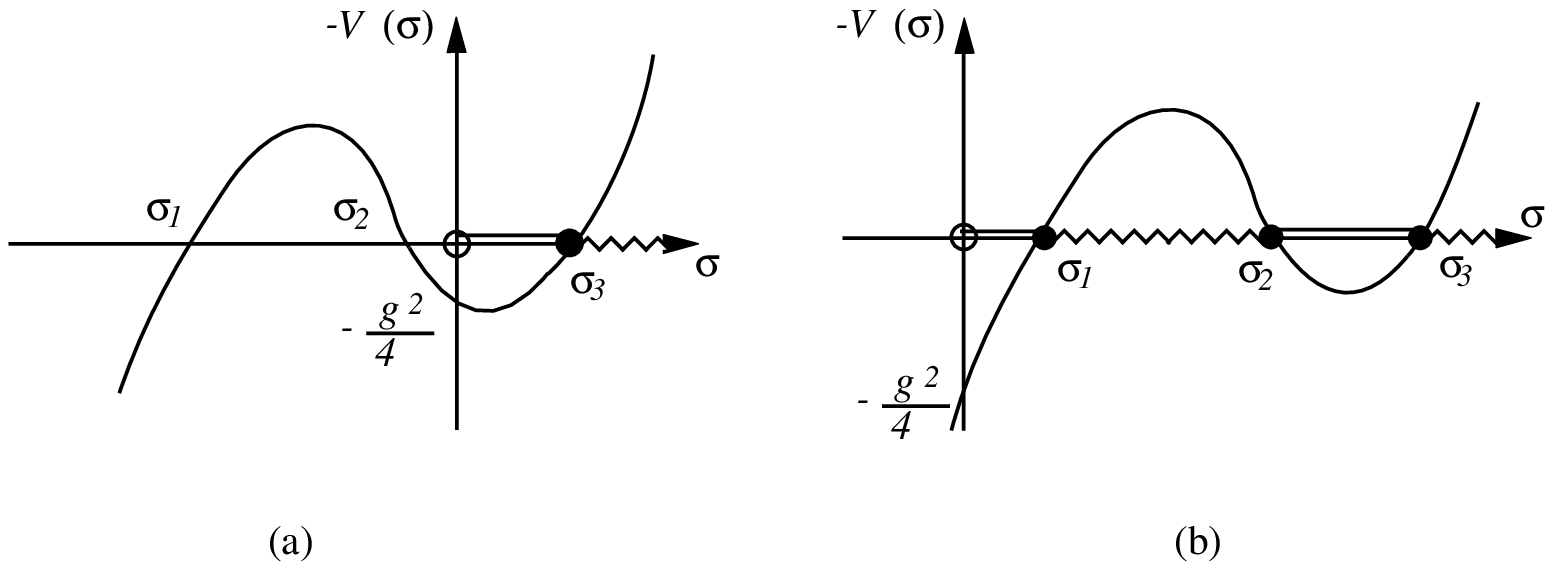}
\end{center}

\vspace{10pt}\nd
FIGURE 2: ~
{The $\beta=-1$ case of Eq. (\ref{2dsiz}).  Either
$\si_1<\si_2<0<\si_3$ (a) or
$0<\si_1<\si_2<\si_3$ (b).  For various notations see caption to Fig. (1).}

\pagebreak

The Schr\"odinger hamiltonian corresponding to the action of Eq. (\ref{2una})
is
\beq
H=-\frac{\hbar^2}{2}\, \Delta+{\cal U}(r)   \label{2ben}
\eeq
where $r^2=\ex^2$, $\Del$ is the $N$ dimensional laplacian and
\beq
{\cal U}(r)= \frac{m^2}{2}\, r^2+  \frac{\beta g^2}{4N}\, r^4\,.\label{2buna}
\eeq

Separating out the radial Schr\"odinger equation for a state carrying angular
momentum $\ell$ we find
\beqra
&& \left\{-\frac{\hbar^2}{2}\left[\frac{\pa^2}{\pa r^2}
+\frac{N-1}{r}\;\frac{\pa}{\pa r}- \frac{\ell(\ell+N-2)}{r^2}\right]+\cu
(r)\right\} R_{n,\ell}(r)=\ce_{n,\ell}\,R_{n,\ell}(r); \nonumber \\
&& R_{n,\ell}(r)\mathop\sim\limits_{r\rta 0}r^{\ell}  \label{2bdos}
\eeqra
where $n$ is the principal quantum number.  Following \cite{witt} we perform
the similarity transformation
\beq
H\rightarrow\til H =r^{-\nu}\, H\,r^\nu,~R_{n,\ell}(r)\rightarrow \chi_{n,\ell}
(r)=r^{-\nu}R_{n,\ell}(r),~\nu=-\frac{N-1}{2} \label{2btre}
\eeq
which disposes of the $\frac{\pa}{\pa r}$ term in Eq.
(\ref{2bdos}).\footnote{This transformation if familiar from elementary quantum
mechanics.  What it does, is essentially swallowing a square root of the radial
jacobian (the
$\rho^{\frac{N-1}{2}}$ factor) into a wave function in an inner product:
$R_{n\ell}(r)\rta\chi_{n,\ell}(r)$, inducing the  corresponding transformation
Eq. (\ref{2btre}) on H.} We finally obtain
\beq
\til{H} \eqv Nh=N\left[-\frac{\hbar^2}{2N^2}\,\frac{\pa^2}{\pa \rho^2} +
\cv(\rho) +w_{n,\ell}(\rho)\right]  \label{2bqrt}
\eeq
where $\rho^2=\ys=r^2/N$ and
\beqra
\cv(\rho) &=& \frac{\hbar^2}{8\rho2}
+\cu(\rho)=\frac{\hbar^2}{8\rho^2}+\frac{m^2}{2}\,\rho^2+
 \frac{\beta g^2}{4}\,\rho^4
\nonumber \\
w_{N,\ell}(\rho) &=& \hbar^2\left[ \frac{\ell-1}{2N} +
\frac{(\ell-\haf)(\ell-\frac{3}{2})}{2N^2}\right]\frac{1}{\rho^2}\,.
\label{2bcin}
\eeqra
Eqs. (\ref{2bdos}) and (\ref{2btre}) imply
\beq
\chi_{n,\ell}(\rho)\mathop\sim\limits_{\rho\rta 0}\rho^{\ell+\frac{N-1}{2}}
\label{2bsiz}
\eeq
assuring the self adjointness of $\til H$.\footnote{Alternatively, Eq.
(2.26)  may be derived directly from Eqs. (\ref{2bqrt})-(\ref{2bcin}).}
For a later reference let us recall that the {\it singular} solution
$\til\chi_{n\ell}$ to the equation $\til H
\til\chi_{n,\ell}=\ce_{n,\ell}\,\til\chi_{n,\ell}$ that is definitely not in
the Hilbert space at hand, blows up at $\rho=0$ as\footnote{This can be seen
immediately from the fact that the wronskian
$W(\chi,\til\chi)$ is necessarily a non vanishing constant.}
\beq
\til\chi_{n,\ell(\rho)}\mathop\sim\limits_{\rho\rta
0}\rho^{1-\ell-\frac{N-1}{2}}\,.  \label{2bset}
\eeq

The explicit $\hbar$ dependence of $\cv$ and $w_{N,\ell}$ in Eq. (\ref{2bcin})
is obvious- its origins are the centrifugal barrier in Eq. (\ref{2bdos}) and
the  similarity transformation we performed on the operator $H$.  To leading
order in $1/N$ the only $\hbar$ dependent term that survives there is the
$\frac{1}{8\rho^2}$ piece in $\cv$.  This is consistent with Langer's
modification of the centrifugal term
$\haf\ell(\ell+N-2)\rta\haf(\ell+\frac{N-2}{2})^2$ in the limit $N\rta \infty$
\cite{love}.
It is therefore the leading
$1/N$ {\it quantum} contribution to the effective potential.  Thus, it must be
equivalent  to the leading $1/N$ correction to the action from the determinant
in Eq. (\ref{2nev}).  We show now that this is indeed the case, namely, as we
have stated earlier the solutions of Eq. (\ref{2dsiz}) correspond exactly to
{\it
classical} motions in the potential
$\cv(\rho)$.  From now on we set $\hbar=1$ for convenience in eqs.
(\ref{2bqrt})-(\ref{2bcin}).  Clearly, from the point of view of the large $N$
approximation, $1/N$ will play the role of a small redefined Planck constant.
Therefore $1/N$ analysis of Eq.(\ref{2bqrt}) is equivalent to its JWKB
analysis.  But this is clearly not the ordinary case considered in a JWKB
approximation, namely, the potential in Eq.(\ref{2bqrt}) depends explicitly on
the small parameter
$1/N$.  While this may complicate the $1/N$ expansion of
the actual {\it eigenvalues} of $h$ in Eq.(\ref{2bqrt}), as far as the $1/N$
expansion of wavefunctions {\it away from the turning points} is concerned,
the situation is much simpler.  Indeed, substituting the expansion
\beq
\chi_\ell(\rho) =\exp\left\{i\,N\left[\sum_{n=0}^\infty
N^{-n}\vphi_n(\rho)\right]\right\}   \label{2boch}
\eeq
into the eigenvalue equation for $h$ in Eq.(\ref{2bqrt}) (assuming the
eigenvalue is independent of $N$ up to an overall scaling) yields to first
subleading order in
$1/N$
\beqra
&&\chi_\ell(\rho) = \left[2\left(\ce-\cv(\rho)\right)\right]^{-1/4}\nonumber \\
&&\exp\left\{\pm iN\int^\rho \left[ 2\left(
\ce-\cv(\rho)\right)\right]^{1/2}\left(1-\frac{\ell-1}{2N\rho^2}\,~
\frac{1}{2(\ce-\cv)} + \cO (N^{-2})\right)d\rho\right\}\,.~~~\label{2bnev}
\eeqra

Therefore, the classical momentum associated with Eq.(\ref{2bnev}) is

\beqra
p(\rho) &=& \left[2\left(\ce-\cv(\rho)\right)\right]^{1/2}\left(1-
\frac{\ell-1}{2N\rho^2}~\frac{1}{2(\ce-\cv)}  +\cO( N^{-2})\right) \nonumber \\
&=&\left[2\left(\ce-\cv(\rho)-w_{N,\ell}(\rho)\right)\right]^{1/2}+
\cO(N^{-2})\,.    \label{2tnta}
\eeqra
Thus, to leading order in $1/N$ we may safely throw away all non leading terms
in the potential   of Eq.(\ref{2bqrt}) (which include all $\ell$ dependent
terms) obtaining
\beq
\frac{p^2}{2}\eqv\haf\,
\al\left(\frac{d\rho}{dt}\right)^2=\ce-\cv(\rho)~~,~~\rho(T)=\rho(0)
\label{2tana}
\eeq
Using Eqs.(\ref{2cin}), (\ref{2bcin}) we find that Eq. (\ref{2tana}) may be
written as
\beq
\al\left(\frac{d\si}{dt}\right)^2 =\frac{g^2}{4}\left(\frac{16\ce}{g}\,
\si-1\right)-4(m^2\si^2+\beta g\si^3),~\si(T)=\si(0)  \label{2tdos}
\eeq
which is nothing but Eq. (\ref{2dsiz}) upon the identification
\beq
\frac{16\ce}{g} = \frac{1}{\si_0}\,.  \label{2ttre}
\eeq

Thus, we have proven
that our calculation of time dependent saddle point configurations $\si(t)$
(Eqs.(\ref{2dqrt})-(\ref{2dsiz})) of $\sef$ is equivalent {\it completely} to
the time dependent classical solutions in the effective potential $\cv$ (Eqs.
(\ref{2bcin})-(\ref{2tana})).  Note  from Eqs. (\ref{2bqrt})-(\ref{2bcin})
that angular momentum effects will show up in $\sef$ only in higher terms of
the
$1/N$ expansion.
 Post factum the equivalence of Eqs. (\ref{2dsiz}) and (\ref{2tana}) seems self
evident, but we have presented an explicit proof of it.  Moreover, in passing
from Eq.(\ref{2tana}) to (\ref{2tdos}) we have used the classical $\si$
equation
of motion (Eq.(\ref{2cin})).  This equation must hold also quantum mechnaically
because we can enforce it as a constraint by introducing a lagrange multiplier
field in an equivalent formulation.  Indeed, note that Eq.(\ref{2cin}) used in
passing from Eq.(\ref{2tana}) to (\ref{2tdos}) relates the {\it slow} radial
semiclassical motions
$\rho(t)$ of $\y$ in the effective potential $\cv$ (that already includes
quantal corrections from $\y$ fluctuations) to the $\si$ fields.

The equivalence of Eqs. (\ref{2dqrt}) and (\ref{2bnev}) holds for their {\it
time  independent} solutions as well.  Indeed, from Eq. (\ref{2bqrt}) the
extremum condition for
$\cv(\rho)$ reads
\beq
\frac{\pa\cv}{\pa \rho} =- \frac{1}{4\rho^3}+m^2\rho+\beta g^2\;\rho^3=0
\label{2tqrt}
\eeq
which is equivalent to Eq. (\ref{2doce}), the equation determining constant
$\si=\si_c$ extremal configurations of $\sef$.  The ground state of the
unharmonic oscillator correspond to the local minimum
 of $\cv(\rho)$.  For $\beta=-1$ this is only a metastable ground state.  It
will decay at a rate suppressed by a factor of
$e^{-N}$ to $-\infty$.  $\cv(\rho)$ has been drawn schematically in figures
3(a)
and 3(b) (for $\beta=1$ and $\beta=-1$, respectively) where we have also
denoted
the various regions along the $\rho$ axis accessible to motions either in real
or imaginary times for some energy parameter $\ce$.  $\ce_0=\cv(\rho_0)$ in
Fig.
(3a) (Fig. (3b)) corresponds to the absolute (metastable) ground state of the
oscillator, thus, using Eqs.
(\ref{2cin}), (\ref{2tqrt}) we have
\beq
\si_c=\haf\, g\,\rho_0^2\;.   \label{2tcin}
\eeq
For $\beta=-1$ Eqs. (\ref{2doce}), (\ref{2tqrt}) have yet another solution,
corresponding to the local {\it maximum} in fig. (3b).

\vspace{-6pt}
\begin{center}
\epsfxsize=5.5in\hspace*{0in}
\epsffile{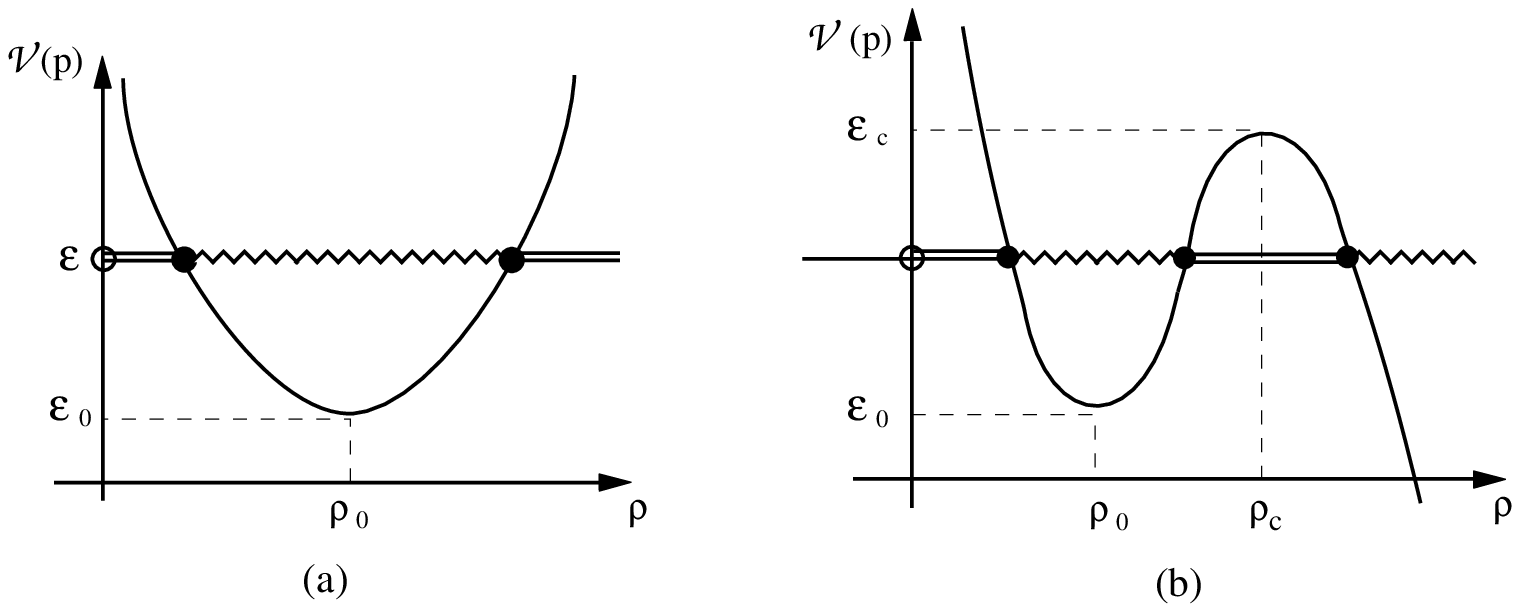}
\end{center}

FIGURE 3: ~
{  Radial motions from Eqs. (\ref{2bqrt}) and (\ref{2tana}) as $N\rta
\infty$ for
$\beta=+1$ (a) and $\beta=-1$ (b).  For the various notions see the caption to
Fig. (1).}

The consistency of Figs. (1a,b) and (3a) and of Figs. (2a,b) and (3b) is
apparent.  Clearly Fig. (1b) corresponds to the case $\ce>\ce_0$ in Fig. (3a)
while Fig. (1a) corresponds to $\ce<\ce_0$.\footnote{With the exception that
$\si_1,\si_2$ there may become complex (conjugate).}  Similarly, Fig. (2a)
corresponds to the cases $\ce>\ce_c$ or $\ce<\ce_0$ in Fig. (3b) while Fig.
(2b) corresponds to $\ce_0<\ce<\ce_c$.\rf{19}

Having figure (3) at hand it is clear now how the turning point structure of
Eq.
(\ref{2dsiz}) and figures (1) and (2) change under variations of the parameter
$\si_0$ holding $m$ and $g$ fixed.  Consider a classical trajectory in Fig.
(3a)
with a very large and positive energy parameter $\ce$.  By virtue of Eq.
(\ref{2ttre}) this corresponds to a very small positive $\si_0$ parameter in
Fig. (1b).  Clearly in this situation we have $0+\sim\si_2\ll\si_3$.  Reducing
$\ce$ in Fig. (3a) (equivalently, increasing $\si_0$), there are no
qualitative changes in Fig. (1b).  It is clear that as $\ce$ decreases,
$\si_2$ increases while
$\si_3$ decreases, diminishing the amplitude of the real-time motion, and in
addition, reducing the value of $-V(\si)$ at its maximum $A$.  These changes
persist until $\ce$ hits $\ce_0=\cv(\rho_0)$, where $\si_2$ and $\si_3$
coincide, shrinking the real time trajectory to a point, corresponding to the
fact that the oscillator is frozen at its ground state.  In this case the
common value of $\si_2$ and $\si_3$ is the constant configuration $\si=\si_c$
mentioned above.  Decreasing
$\ce$ further, we flip from Fig (1b) to Fig. (1a), where only imaginary time
motions are allowed.  A similar behaviour occurs in figures (3b) and (2a,b).

Figures (1)-(3) exhibit imaginary time solutions to Eqs. (\ref{2dsiz}) and
(\ref{2tana}) that hit the boundary point $\si=0$ for any value of $\ce$ or
$\si_0$.  Such solutions describe  situations in which the particle may be
close to the end-point $\si=0$ during a finite time period.  This is in obvious
contradiction with the small $\rho$ behavior of the square integrable wave
functions $\chi_{n,\ell}(\rho)$ in Eq. (\ref{2bsiz}) which suppress the
probability of having the oscillator oscillating at $0\sim\rho\ll 1$ {\it
completely} (for any
$\ell$) as $N\rta \infty$.\footnote{The same problem arises for a {\it free}
particle in $N$ dimensions and imaginary time.}\rf,\footnote{It might be though
that these classical trajectories are associated with the functions
$\til\chi_{n,\ell}$ in Eq. (\ref{2bset}) that are solutions of the
Schr\"odinger
equation, but do not belong to the Hilbert space.}  We will not discuss these
solutions further.

As was mentioned at the beginning of this section, a flip in the sign of $m^2$
is equivalent to a simultaneous flip of $\al$ and $\beta$, i.e. a change from
real (imaginary) time motions in Fig. (3a) to imaginary (real) time motions in
Fig. (3b).  Therefore, if one starts with $m^2>0$, a flip in the sign of $m^2$
will not induce a local minimum in addition to the one that already exists in
Figs. (3a) and (3b) at $\rho=\rho_0$.  Alternatively, one can check explicitly,
that the local minimum of $\cu(\rho)$ in Eq. (\ref{2buna}) at $\rho=-m^2/g^2$
(for $m^2<0$) is distorted away by the $1/8\rho^2$ term in in Eq.
(\ref{2bcin}).

 We end this section with some
general observations and remarks on the possible solutions $\si(t)$ to Eq.
(\ref{2dsiz}).  Following our discussion of the various possible trajectories,
a
straightforward integration of Eq. (\ref{2dsiz}) yields the latter as
\beq
\si(t) =\psi(t-t_0)  \label{2tsiz}
\eeq
where $\psi$ is an elliptic function \cite{ww,ast} and $t_0$ is an integration
constant assuring time translational invariance.  As should be clear from Figs.
(1)-(3) the amplitudes of the oscillator are either bounded (the poles of
$\psi$ in Eq.(\ref{2tsiz}) are away from the real $t$ axis) or  unbounded
($\psi$ has a double pole on the real
$t$ axis).  In the latter case $\psi(t-t_0)$ is essentially a Weierstrass $\cp$
function, while in the former case, it may be written most simply as a
rational expression in terms of the Jacobi $sn$ function.  These finite
amplitude solutions descend into the expected harmonic oscillations in real
(imaginary) time as $\ce\rta\ce_0+$ ($\ce\rta\ce_c-$) in Figs. (3a) and (3b),
where the oscillator executes small oscillations around the relevant minimum.
The unbounded oscillations on the other hand are potential sources to an
infinite $\sef$, which can suppress such modes completely. Clearly, as they
stand, they are not even periodic in time, since there is nothing to reflect
$\si(t)$ back from infinity.  Moreover, these modes are associated with very
large values of
$\ex$, which might throw us out of the validity domain of the large $N$
approximation.  However, regardless of all these problems, the Weierstrass
$\cp$ function
$\si(t)$ configurations are very interesting since for such $\si$'s the
Schr\"odinger hamiltonian $h$ in Eq.(\ref{2dtre}), which is essentially the
inverse propagator of the $\y$ field in Eq. (\ref{2qrto}), is one of the
completely integrable one dimensional hamiltonians of the Calogero type
\cite{calogero}.

For real time motions, such trajectories occur {\it only} for $\beta=-1$ (see
Fig. (3)), i.e. only in the case of negative quartic coupling.  In such a case
the system, strictly speaking, does not have a ground state and is meaningless
beyond large $N$ saddle point considerations.  However, due to the very fast
decrease of the potential to $-\infty$, the time of flight to infinity is
finite, and the unharmonic oscillator attains in this case a one parameter self
adjoint extension \cite{selad} that might be used to redefine the quantal
system into a well behaved form that will render these interesting $\cp$
function shape trajectories meaningful.  In this way we actually compactify
the $\si$ (or $\rho$) semiaxis, making these trajectories effectively
periodic, in accordance with the periodicity of the $\cp$ function.

Any of the semiclassical trajectories $\rho(t)$ with bounded amplitude from
fig. (3) or their corresponding $\si(t)$ configurations in figs. (1)  and (2)
defines a vibrational mode of the $\ON$ vector.  In such a vibrational mode,
the
unharmonic oscillator is described effectively by an harmonic oscillator that
is coupled to a time dependent potential $v(t)=\frac{m^2}{2} + \beta\si(t)$
(Eq. (\ref{2qrto})).  Each vibrational mode corresponds to a unique $\si(t)$
configuration found from Eqs. (\ref{2dsiz}) or (\ref{2tdos}).  On top of these
vibrations, the $\ON$ vector rotates as well, giving rise to a host of
rotational states that form a band superimposed on top of the vibrational mode.
To leading order in $1/N$, this rotational band collapses into a single energy
vlaue as should be clear from Eq. (\ref{2bcin}).

Each vibrational mode $\si(t)$ is an {\it unstable} saddle point of $\sef$.
Indeed it can be shown straightforwardly that the second variation of $\sef$
around each such $\si(t)$ configuration has an infinite number of negative
directions, corresponding to the infinite number of decay modes of this excited
mode as $N\rta\infty$.  Nevertheless, we must include these unstable saddle
point configurations in calculating $W_p(T)$ in Eq. (\ref{2och}) for finite
$T$,
since excited states of the oscillator do contribute to it.  We will not pursue
the case of the unharmonic oscillator any further in this paper, but rather
turn
to the two dimensional field theoretic case in the next section.

\pagebreak

\section{The ${\cal O}(N)$ Vector Model in Two
Dimensions}
\setcounter{equation}{0}
\renewcommand{\theequation}{3.\arabic{equation}}

\indent\indent The action for the two dimensional $\ON$ model reads
\beq
S=\int\limits_0^T dt \int\limits^\infty_{-\infty} dx\,
\left[\haf\left(\pa_\mu\,\bPh\right)^2-\frac{\al m^2}{2}\;\bPhs -
\frac{\al\beta g^2}{4N} \left(\bPhs\right)^2\right] \label{3una}
\eeq
where $\bPh$ is an $N$ component scalar field transforming as an $\ON$
vector,
$m^2$ is the bare ``meson'' mass and $g^2$ is the {\it finite} quartic
coupling.
$\al,\beta$ and $\gamma$ (that appears below) have the same meaning as in
the previous section, hence
$(\pa_\mu\bPh)^2=(\pa_0\bPh)^2-\al(\pa_1\bPh)^2$.
Rescaling
\beq
\bPhs=N \bphs         \label{3dos}
\eeq
and introducing the auxiliary field $\sigx$ as before, Eq. (\ref{3una})
turns into
\beq
S=N\int^T_0 dt \int^\infty_{-\infty} dx
\,\left[\haf(\pa_\mu\bph)^2-\frac{\al m^2}{2}\, \bphs + \ab
(\si^2-g\si\bphs)\right]\;. \label{3tres}
\eeq
Note again, that by its equations of motion $\si=\frac{g}{2}\,\bphs$ is a
non-negative field (recall our convention $g\geq 0$).  This positivity
should persist in the quantal domain as well, since this equation of motion
may be enforced equivalently as a constraint by introducing a Lagrange
multiplier.

In complete analogy with Eqs. (\ref{2siz})-(\ref{2nev}) we define the
amplitude
\beqra
W_P(T) &=&\lag \bph_0~;~ t=T \bmid \bph_0~;~t=0 \rag = \nonumber\\[5pt]
&=& \int \cd_{pbc}\si\; e^{N\gamma\sef} \label{3qrto}
\eeqra
where $\bph$ has been integrated out yielding formally
\beq
\sef=\ab\int\limits_0^T dt\int\limits_{-\infty}^\infty dx
\si^2-\frac{1}{2\gamma}\tr_{pbc}\ln\left[- \bx-\al m^2-2\ab g\si\right]
\label{3cin}
\eeq
in which $\bx =\pa_0^2-\al\pa_1^2$.

As was mentioned in the introduction the ground state of the $\ON$ model
is governed by the homogeneous constant $\si=\si_c$ configuration, where
$\si_c$ found from by the extremum condition
$\displaystyle{\left.\frac{\del\sef}{\del\sigx}\right|_{\si=\si_c}=0}$,
the so called ``gap equation'' of the model,  and is a {\it local} minimum
of $\sef$.  This gap equation expressed in terms of the bare quantities $m$
and $g$ in Eq. (\ref{3una}) contains logarithmic divergences that can be
swallowed into the bare
$m^2$ parameter, replacing it by a renormalized finite $m^2_R$ parameter,
that depends on an arbitrary finite mass scale, while
$g^2$ remains unchanged \cite{larn,sh,dvc,mdvc}.

In general, this ``gap equation'' might have more than one solution
$\si=\si_c$, but only one such
solution corresponds to the ground state of the theory.  The latter is
$\ON$ invariant \cite{sh} as we are discussing a two dimensional field
theory with a continuous symmetry \cite{cmw}.

The local minimum $\si=\si_c$ of $\sef$ supports small fluctuations of the
fields in Eq. (\ref{3tres}) around it giving rise to a low energy spectrum
of ``free mesons" with mass squared $m^2_R+2\beta  g\si_c$.  There exist
also bound states and resonances in the $\ON$ singlet $\bph\cdot\bph$
channel
\cite{sh}.  This picture is intuitively clear in the case of positive
quartic interactions
$\beta=1$, but it turns out to persist also in the $\beta=-1$ case of
negative quartic coupling \cite{sh}, provided $g^2$ is not too large.
What happens in the latter case is that despite the fact that for
$\beta=-1$  Eq. (\ref{3una}) has no ground state, $\sef$ still possesses a
local minimum that support small oscillations of the fields in Eq.
(\ref{3una}).  To leading order in $1/N$ these fluctuations cannot spill
over the adjacent local maximum of
$\sef$ towards infinite field values, provided $g^2$ does not exceed a
certain maximal critical value.  It was found in \cite{sh} that the
$\si-\si$  propagator is indeed free of tachyons even in the $\beta=-1$
case, provided
$g^2$ is smaller than that critical bound.

We now turn to the central topic in this work, namely, the static $\sigx$
extremal  configurations of $\sef$, and the bands of bound $\bph$ states
associated with them.  Note that such static $\sigx$ configurations satisfy the
periodic boundary condition along the time direction

\beq
\si(x,0)=\si(x,T) \label{3siz}
\eeq
implied by Eq. (\ref{3qrto}) automatically, for any value of $T$.

Our approach to these static $\sigx$ configurations will be indirect.
Namely, we assume that the extremal $\si(x,t)$ configuration we are looking
for has a {\it very slow} time variation, such that $[\pa_0,\si]\approx 0$
will be a very good approximation in calculating the trace in Eq.
(\ref{3cin}). In this ``adiabatic'' approximation we effectively carry a
smooth dimensional reduction of the {\it extremum condition}
$\frac{\del\sef}{\del\si(x,t) } =0$ from $1+1$ to $1+0$ space-time
dimensions.  Therefore, we use this ``adiabatic'' limit merely as an
infra-red regulator of the $t$-integrations that occur in matrix elements
in the trace of Eq. (\ref{3cin}) which enables us to extract the static
$\sigx=\si_0(x)$ piece.  We do not know if the slowly time varying
$\si(x,t)$ configurations we encounter in this approximation are relevant
as  genuine space-time dependent fields, or just artifacts of the
approximation.

Carrying the ``adiabatic'' approximation we expand the real field $\si$ in
frequency modes

\beq
\si(x,t)=\sum_{n=-\infty}^\infty \si_n(x)e^{i\om_nt}  \label{3set}
\eeq
where by virtue of Eq. (\ref{3siz}) and the fact that $\si^*=\si$ we have

\beq
\om_n=\frac{2\pi}{T}\,P(n)~;~P(-n)=-P(n)~;~\si_n^*(x) =\si_{-n}(x)
\label{3och}
\eeq
in which $P(n)$ is {\it any} monotonously increasing map from the integers
to themselves.  The precise form of $P(n)$ will be immaterial to us, since
we are interested only in the zero frequency mode $\si_0(x)$ in Eq.
(\ref{3set}).

Using Eqs. (\ref{3set}) and (\ref{3och}) $\sef$ in Eq. (\ref{3cin})
becomes
\beqra
\sef &=& \ab\, T\int\limits_{-\infty}^\infty
dx\,\left[\si_0^2(x)+2\sum_{n=1}^\infty\,\si_{-n}(x)\si_n(x)\right]
\nonumber \\[5pt]
&&-\frac{1}{2\gamma}\int\limits^\infty_{-\infty} dx\sum_{n=-\infty}^\infty
\lag x,n|\ln\left[\om_n^2+\al\pax^2-\al m^2-2\ab g\si_n\right]\,|x,n\rag
\label{3nev}
\eeqra
where $\left\{|x,n\rag\right\}$ is a complete orthonormal basis of
positions
$x$ and frequency modes $\om_n$ and we have used the adiabatic
approximation
$\left[\pa_0,\si\right]
\approx 0$.

Varying Eq. (\ref{3nev}) with respect to $\si_{-k}\;(k\geq 0)$ the
extremum condition reads

\beq
\si_k(x)=\frac{g}{2\al\gam T}\lag x|\, \frac{1}{-\pax^2+2\beta g\si_{-k}
+ m^2-\al\om_k^2}\, |x\rag \label{3dez}
\eeq
for the $k$-th mode in Eq.  (\ref{3set}).  Here we have used the relations
$\lag n|n\rag=1$ and $\al^2=1$.  Eq.  (\ref{3dez}) is of the general form
of Eq.  (\ref{2onte}) and thus shows explicitly that our adiabatic
approximation has reduced the $1+1$ dimensional extremum condition into a
$0+1$ dimensional problem, as was stated above.  Indeed, Eq.  (\ref{3dez})
for $k=0$ could have been obtained equivalently by ignoring completely any
time dependence in the extremum condition

\beq
\frac{\delta\sef}{\del\si(x,t)} = \ab\,\left[ 2\si(x,t) + \frac{g}{\gam}\,
\lag x,t|\, \frac{1}{-\pa_0^2+\al\pa_1^2-\al m^2-2\ab g\si}\,
|x,t\rag\right]=0,
\label{3onte}
\eeq
namely discarding $t,\,|t\rag$ and $\pa_0$ in this equation from the very
beginning.  In order to maintain the correct scale dimensions of the
remaining objects, one simply replaces $g$ by $g/T$ in Eq.  (\ref
{3onte}),  obtaining Eq.  (\ref {3dez}) again.

The advantage in performing the adiabatic approximation (Eqs.
 (\ref{3set})-(\ref{3dez})\,) rather than the simpler derivation of the
extremum condition for $\si_0(x)$ we have just describe is that it
enables us  to see explicitly what happens to non-zero frequency modes in
Eqs.
 (\ref {3set})- (\ref{3dez}).

At this point it is very important to note that we must {\it exclude} space
independent solutions to  Eq.  (\ref{3dez}) (as far as $k=0$ is concerned).
For such solutions  Eq.  (\ref {3dez}) is equivalent to an {\it algebraic
cubic} equation similar to  Eq.  (\ref{2doce}).  However, the homogeneous
$\si=\si_c$ vacuum condensate satisfies the ``gap-equation'' which is a
transcendental equation \cite{sh,abbott} rather than an algebraic one.  For
this reason it is very clear that our resulting static $\si_0(x)$
configuration will be very different from the one found in \cite{abbott},
which is only a {\it mild} distortion of the vacuum $\si_c$ condensate.
Thus, the band of bound states associated with our
$\si_0(x)$ complements those discussed in \cite{abbott}.

Since we are interested essentially in the zero frequency mode of  Eq.
 (\ref {3set}), we are free to assume that $\si(x,t)$ is an even function
of time,  namely

\beq
\si_{-k}(x) = \si_k(x) = \si_k^* (x)   \label{3doce}
\eeq

In this case, we see that  Eq.  (\ref {3dez}) relates the {\it diagonal
resolvent} $R$ of the Schr\"odinger operator

\beq
h=-\pax^2+2\beta g\si_k(x) \label{3tre}
\eeq
at energy parameter $\al\om_k^2-m^2$, to its potential term $\si_k(x)$:

\beq
R(x)=\frac{2\al\gam T}{g}\, \si_k(x)\;, \label{3dqrt}
\eeq
in complete analogy with  Eq.  (\ref{2onte}).

Following the same steps as in the previous section, we find from the
Gelfand-Dikii equation (Eq. A.12) that $\si_k$ must satisfy the equation

\beq
2\si_k\si_k^{\prime\prime}-\left(\si_k'\right)^2 - 4\, (2\beta g\si_k +
m^2-\al\om^2_k)\,\si_k^2 =\al\left(\frac{g}{2T}\right)^2 \label{3dcin}
\eeq

\vspace{8pt} \nd
where we have used the relations $\al^2=1$ and $\gam^2=-\al$.
Note from  Eq.  (\ref{3dez}) that  Eq.  (\ref{3dcin}) holds for the zero
frequency mode $\si_0(x)$ independently of whether the adiabatic extension
$\si(x,t)$ of the static configuration has any definite parity properties
under
$t\rta-t$ or not.

Naturally,  Eq.  (\ref {3dsiz}) coincides with  Eq.  (\ref {2dqrt}) in the
case of euclidean signature $\al=-1$ upon the replacements $m^2\rta m^2 +
\om^2$,
$g\rta g/T$.

As in the previous section, first integration of  Eq.  (\ref {3dcin}) is
done by substituting

\beq
\si'_k(x)=f_k(\si_k) \label{3dsiz}
\eeq

\vspace{8pt} \nd
leading to

\beq
f_k^2=\left(\frac{d\si_k}{dx}\right)^2=4\left[ \beta
g\si_k^3+\left(m^2-\al\om_k^2\right)\si^2\right] +
\al\left(\frac{g}{2T}\right)^2
\left(\frac{\si_k}{\til\si_k}-1\right) \label{3dset}
\eeq
\vspace{8pt} \nd
where $\til\si$ is an integration constant.  Note that unlike the $0+1$
dimensional model discussed in the previous section, in the field
theoretic  case we lack a simple physical
interpretation of  Eq.   (\ref {3dset})  as an equation of motion in one
dimension allowing us to discard ``radial'' trajectories (in euclidean
signature) that hit the origin
$\si=0$.

Simple analysis of  Eq.  (\ref {3dset}) for very high frequencies $\om_k$,
reveals that the corresponding modes $\si_k(x)$ with bounded amplitudes are
highly suppressed.  Namely, one finds that $\si_k$ is of the order
$\frac{g^2}{\til\si(\om_kT)^2} \ll 1$.  Therefore, if we {\it choose}
$P(n)$ in  Eq.  (\ref {3och}) such that already $2\pi \, P(1) \gg
1$~ (e.g.
$P(n)=ne^{(n^2+1)^{1000}}$), then for large enough $T$ and generic values
of
$\til\si$ {\it all} non zero frequency modes will be highly suppressed in
the resulting extremum condition, assuring self consistency of the
adiabatic approximation.  As $\si_0(x)$ is decoupled from non zero
frequency modes within the  framework of the approximation made, we are
free to make such choices of the $P(n)$'s.

Concentrating on the zero-frequency mode,  Eq.  (\ref{3dset}) becomes

\beq
\left(\frac{d\si}{dx}\right)^2 \eqv f^2(\si)=4\left[\beta g\si^3 +
m^2\si^2\right] +\al \left(\frac{g}{2T}\right)\,\left(\frac{\si}{\til\si}
-1\right)     \label{3doch}
\eeq

\vspace{8pt} \nd
where we have dropped the sub-index $k=0$ everywhere.  Note the explicit
$T$ dependence of the right hand side of  Eq.  (\ref{3doch}).
This implies that $\si_0(x)$, which is our {\it time-independent} saddle
point configuration, will have a peculiar dependence on $T$.

However, this paradox will be resolved by the fact that solutions to  Eq.
 (\ref{3doch}) in which the $T$ dependence is ``important" (in a sense
that will become clear when we consider these solutions) always give rise
to {\it infinite} $\sef$ values, and are thus highly suppressed in the path
integral ( Eq.  (\ref{3qrto}), unless we put the field theory in a finite
``spatial'' box of length $L$.  In the latter case, these $T$ dependent
$\si_0(x)$ solutions are {\it not} suppressed. In such cases
having explicit $T$ dependence is consistent with having explicit $L$
dependence.  We will not discuss the interesting finite size behavior of
this field theory and its corresponding thermodynamic behaviour in this
paper.

Thus, if $\si_1,\,\si_2$ and $\si_3$ are the three roots of the cubic
equation $f^2(0)=0$, we have

\begin{subeqnarray}
f^2(0) &=& -\al\left(\frac{g}{2T}\right)^2 \label{3dnev1} \\[4pt]
f^2(\pm\infty) &=& \pm \beta\,\infty    \label{3dnev2} \\[4pt]
\si_1\si_2\si_3 &=& \frac{\ab g}{16T^2}     \label{3dnev3} \\[4pt]
\si_1+\si_2+\si_3 &=& - \frac{\beta\,m^2}{g}\;.  \label{3dnev4}
\end{subeqnarray}

\vspace{8pt}
Here we also adopt the convention that if $\si_1, \,\si_2$ and $\si_3$
are all real, then \break
$\si_1\leq \si_2\leq \si_3$.  Unlike Eq. (\ref{2dsiz}), $\al$
is not an overall factor in  (\ref{3doch}) and we shall have to analyze
all  four possible $(\al,\beta)$ combinations separately.  We discuss
below only cases in which $\si_1,\si_2$ and $\si_3$ are all {\it real}.

Clearly, then, (\ref{3dnev3}c) implies that each $\al,\beta$ combination
gives rise to two possibilities, according to the arithmetic signs of
$\si_1,\si_2$ and $\si_3$, yielding eight possibilities in all.  These
eight $\si_0(x)$ configurations are depicted schematically in figs.
(4a)-(4d).   However, two of these (figs. (4a.i) and (4b.i) are explicitly
ruled out by (\ref{3dnev4}d), since it implies that when $\si_1,\si_2$ and
$\si_3$ have all the same sign, the latter must be $-\beta$.  In the
remaining configurations of fig. (4), the physically allowed domains for
$\si$ are segments on the semiaxis
$\si>0$ along which $f^2(\si) \geq 0$.

As in the previous section, it is clear from  (\ref{3doch}) and figs.
(4a-d) that all static $\si$ configurations are of the form
$\si(x)=\psi(x-x_0)$   where
$x_0$ is an integration parameter and $\psi$ is either (essentially) a
Weierstrass $\cp$ function (unbounded $\si$ configurations) or a simple
rational function in terms of Jacobi functions, e.g. the {\it sn} or
{\it cn} functions (bounded $\si$ configurations).  All these $\si$
configurations are periodic in $x$.  In euclidean space-time signature we
have again the $\si$ configurations that hit the endpoint $\si=0$.  Unlike
the $0+1$ dimensional case discussed in the previous section, we have no a
priori reason to discard such solutions, but note that as functions of $x$,
their first derivative suffers a finite jump discontinuity whenever $\si$
vanishes.  This fact makes such $\sigx$ configurations rather suspicious.

\pagebreak
\begin{center}
\epsfxsize=5.5in\hspace*{0in}
\epsffile{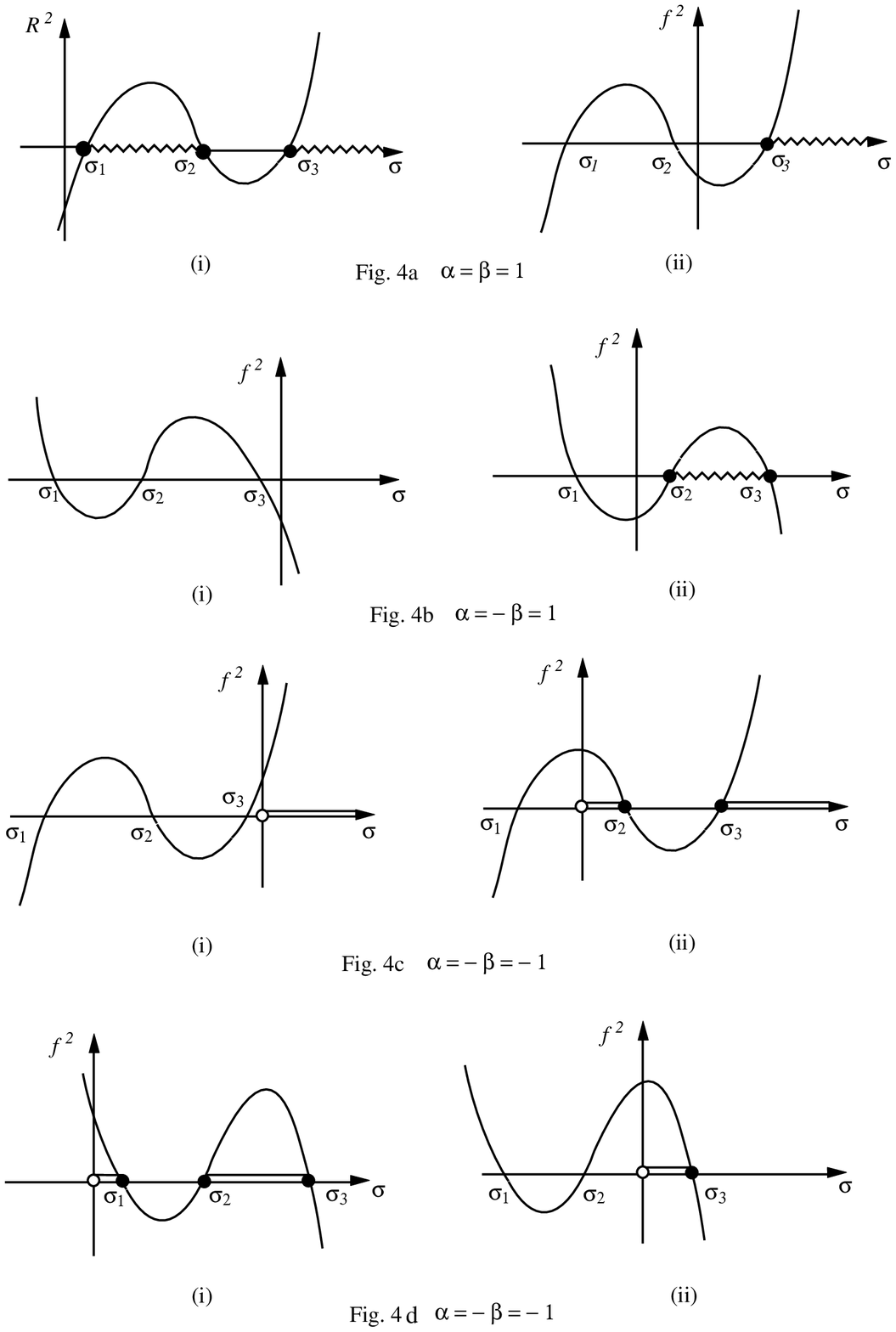}

\vspace{8pt}
FIGURE 4
\end{center}

\pagebreak

We will discuss explicit solutions of Eq. (\ref{3doch}) in the last part
of this section.  At this point we make a small digression to recall the
way the spectrum of ``mesonic" bound states is calculated, given an extremal
static
$\sigx$ configuration (i.e., a specific solution to Eq. (\ref{3doch}) in
our case).  In this digression we follow \cite{dhn3,abbott} very closely.
Given a {\it static} $\si$ configuration and ignoring the gaussian
fluctuations of $\si$ around it, dynamics of the ``meson" $\bph$ field in
this background is governed by the action in Eq. (\ref{3tres}) upon
substituting into it the particular $\sigx$ configuration we are
discussing.   Clearly then the fluctuations of $\bph$ satisfy
\beq
[\bx +\al\,m^2+2\ab g\sigx ]\phi_a=0;\qquad  a=1,\ldots N
\label{3ben}
\eeq
with the periodic boundary conditions
\beq
\phi_a(t+T,\,x)=\phi_a(t,x) \label{3buna}\,.
\eeq
Upon a simple separation of variables $\phi_a(x,t)=e^{i\la t}\phi_a(x)$
Eq. (\ref{3ben}) turns into the eigenvalue problem of the time independent
Schr\"odinger equation\,\footnote{ We have omitted an overall factor $\al$
  from one of the sides  of Eq. (\ref{3buna}).  This will show up later.}
\beq
\left[-\pa^2_x+m^2 + 2\beta g\sigx\right]\phi_a(x) = \la^2[\si]\phi_a(x)
\label{3bdos}
\eeq
where the functional dependence of the eigenvalue $\la^2$ on the
Schr\"odinger potential $\sigx$ has been written explicitly.

Following Eq. (\ref{3bdos})  we expand $\phi_a(x)$ as
\beq
\phi_a(x,t)=\sum_i\, A_a^i(t)\, \psi_i(x)  \label{3btre}
\eeq
where $\{\psi_i(x)\}$ are the complete orthonormal set of eigenfunctions
of the Schr\"odinger hamiltonian in Eq. (\ref{3bdos}) with eigenvalues
$\{\la_i^2\}$, and $A_a^i(t)$ are arbitrary amplitudes to be integrated
over in the functional integral.  The latter are subjected to the periodic
boundary condition
$A_a^i(t+T)= A_a^i(t)$.

Using Eqs. (\ref{3ben})-(\ref{3btre}) the $\bph$ dependent part of Eq.
(\ref{3tres}) becomes
\beq
S= \frac{N}{2}\, \int\limits^T_0 dt\sum_{a=1}^N\,\sum_i\,
\left[(\pa_0A^i_a)^2-\al\la_i^2 (A_a^i)^2\right] \label{3bqrt}
\eeq
which is nothing but the action $N$ identical (infinite) sets of harmonic
oscillators oscillating at frequencies $\{\la_i\}$.

The natural place to look for bound state energies is the partition
function associated with Eq. (\ref{3bqrt}), i.e.  with the particular
static $\sigx$ configuration we are considering.

Using well known results \cite{matned,dhn3,abbott} to integrate over the
$A_a^i(t)$ in Eq. (\ref{3bqrt}), the desired partition function
associated with
$\sigx$
\beq
\cz[\si]= \tr\, e^{H[\si]T/\gam} =\int \cd_{pbc}\bph\; W_P \left(T;
[\si]\right)
\label{3bcin}
\eeq
(where $W_P(T;[\si])$ is the contribution of our particular $\sigx$
configuration to Eq. (\ref{3qrto})\,) may be expressed straightforwardly
as
\beqra
\cz[\si] &=&\left[\prod_i\;\frac{e^{\la_iT/2\gam}}{1-e^{\la_iT/\gam}}
\right]^N\cdot e^{\ab \gam NT \int\limits^\infty_{-\infty} \si^2dx} \nonumber
\\
&=& e^{NT\left[\ab\gam\int^\infty_{-\infty}\si^2dx+\frac{1}{2\gam}\sum_i
\la_i\right]} \sum_{\{n\}} C\left(\{n\}\right) e^{\frac{T}{\gam}\sum_j
n_j\la_j}\;.
\label{3bsiz}
\eeqra

The sum in Eq. (\ref{3bsiz}) runs over all {\it ordered} sets $\{n\}$  of
non negative integers $n_i$ and
\beq
C(\{n\})= \prod_{n_i\eps\{n\}}\;\frac{(N+n_i-1)!}{(N-1)!\;n_i!}\;.
\label{3bset}
\eeq
By definition we have (Eq. (\ref{3bcin}))
\beq
\cz[\si] = \sum_{\{n\}} C(\{n\})\,e^{E(\{n\})T/\gam} \label{3boch}
\eeq
where\footnote{We use the relations $\gam^2=-\al,~~\al^2=1$.}
\beq
E(\{n\}) = \left[-\beta N \int^\infty_{-\infty}\si^2dx+\frac{N}{2} \,
\sum_i
\la_i\right] + \sum_{n_j\eps\{n\}}n_j\la_j \label{3bnev}
\eeq
is the energy of the state corresponding to the set $\{n\}$ \footnote{That
is, inserting $n_k$ ``mesons'' into energy level $\la_k~,~
k=1,2,\ldots$.}$^,$
\footnote{Clearly $E(\{n\})$ corresponds to a ``mesonic'' bound state only
if all quantum numbers $n_k$ that correspond to scattering states in Eq.
(\ref{3bdos}) vanish.} and
$C(\{n\})$ is its degeneracy.

Therefore each $\sigx$ saddle point configuration gives rise to a band of
bound states (as well as scattering states) of $\bph$ ``mesons'' with a very
rich $\ON$ structure.  Generically, the degeneracy factors of energy states
in the band are very large and each $E(\{n\})$ multiplet forms a highly
reducible representation of $\ON$.  The lowest state in a band has no
``mesons'' in it at all ($n_i=0$ for all $i$'s) and its energy is given by
\beq
E_0[\si]=-\beta N\int\limits^\infty_{-\infty}\si^2dx+\frac{N}{2}\;
\sum_i\la_i[\si]
\label{3tnta}
\eeq
where  the first term is the classical energy of the $\sigx$ field and the
other term is the zero point contribution of the oscillators in Eq.
(\ref{3bqrt}).

In general both terms in Eq. (\ref{3tnta}) are divergent and must be
regularized.  In certain cases where the period of $\sigx$ as a function
of $x$ becomes infinite regularization of Eq. (\ref{3tnta}) may be
achieved simply by subtracting from it the zero point contribution of
oscillators in the background of the vacuum configuration $\si=\si_c$,
namely the quantity
$\sum_i\la_i[\si_c]$ \cite{dhn3,abbott}.\footnote{Note in this respect
that our
$2g\si_c$ corresponds to $(\chi-\overline\chi)/N$ in \cite{abbott}.}  We
elaborate on such cases below.  In other cases $E_0[\si]$ might be
inherently divergent, suppressing contributions from such $\sigx$
configurations enormously relative to the vacuum.  Such cases include the
Weierstrass $\cp$ function configurations as well as finite amplitude
$\sigx$'s with a {\it finite} period along the $x$ axis.  However, the
latter may become important if we put the field theory into a spatial
``box'' of finite length $L$, that is an integral multiple of period of
$\sigx$.  As $L\rta\infty$ these modes are suppressed as was stated above.

In \cite{dhn3,abbott} Eqs. (\ref{3bnev}) and (\ref{3tnta}) were the
starting point in a calculation determining extremal $\sigx$
configurations.  In these papers one extremises the regularized form of
Eq. (\ref{3bnev}) with respect to
$\sigx$.  In order to account for the functional dependence of the $\la_i$
on
$\sigx$ one has to invoke inverse scattering techniques.  In these papers,
the
$\sigx$ configurations that resulted from the inverse scattering analysis
were generically very mild distortions of the vacuum condensation
$\si=\si_c$ obtained from the gap equation, expressed  in terms of
reflectionless Schr\"odinger potentials.  Since $\si=\si_c$ is a {\it local
minimum} of the appropriate $\sef$, it is clear from the self consistency
of these analyses that such $\sigx$ configurations must be also local
minima of $\sef$, since they are just the exact result (as $N\rta\infty$)
for the back-reaction of either bosons or fermions on the vacuum, which is
in turn the local minimum of
$\sef$ trapping these particles.

As was mentioned above briefly, our static extremal $\sigx$ configurations
are very different from those found in \cite{dhn3,abbott}.  Moreover,
contrary to the latter, they are generically {\it not} local minima of
$\sef$ but rather saddle points of this object such that the second
variation of $\sef$ around them is not positive definite.  We have seen in
the previous section that despite this fact the contribution of these
saddle points to the path integral representation of the partition function
$\tr\; e^{HT/\gam}$ is important, because they correspond to
(collective) excited states of the  quantized theory.  Clearly, only the
extremal configuration that describes the ground state of the quantum
theory must be a local minimum of
$\sef$. We may strengthen our conclusion of the importance of these $\sigx$
configurations by making an analogy with a simpler elementary example.
Namely, consider a variational calculation of the spectrum of a
Schr\"odinger hamiltonian (for simplicity we consider the one dimensional
case)
\beq
h=-\pa^2_x+V(x)\;.  \label{3tuna}
\eeq
The energy functional is
\beq
S\left[\psi^{\dagger},\psi;E\right] =\int\limits_a^b
\psi^{\dagger}(h-E)\psi\;dx+ E    \label{3tdos}
\eeq
where $E$ is a Lagrange multiplier enforcing the normalization condition
$\lag\psi|\psi\rag=1$ and Neumann boundary conditions are assumed at the
end points.

The extremum condition $\frac{\del S}{\del\psi^{\dagger}(x)} = 0$ reads
\beq
(h-E)\psi=0~~;~~\psi'(a)=\psi'(b)=0\;. \label{3ttre}
\eeq
Obviously, Eq. (\ref{3ttre}) is nothing but the Schr\"odinger equation of
$h$, and thus has an infinite number of solutions
$\{\psi_n(x),E_n\}~~(n=0,1,2\ldots)$ -- the spectrum of $h$.  Clearly, only
{\it
one} extremum is an (absolute) minimum of $S$ - this is the ground state
$\{\psi_0(x),\;E_0\}$.  The other extrema corresponding to excited states
are only saddle points of $H$.  Indeed, the $n$-th state,  $\psi_n(x)$, has
exactly
$n$ unstable directions in its saddle point (recall we are considering the
one dimensional case where no degeneracy occurs),  corresponding to the $n$
states below it.  This is clear from the second variation of $S$ around
$\{\psi_n(x),\; E_n\}$
\beq
S\left[\psi^{\dagger}_n+\del\psi^{\dagger},\;\psi_n+\del\psi,\;E_n\right] =
\sum_{m=0}^\infty\, |C_m|^2(E_m-E_n) + E_n \label{3tqrt}
\eeq
where $\del\psi=\sum^\infty_{m=0} C_m\psi_m(x)$ and we have used the
orthonormality of the $\{\psi_n\}$.  These saddle points are important
as they correspond to all excited states of the spectrum and must be taken
into account.

Having established the importance of our static saddle point
 configurations we now turn to the explicit calculation of these.  Out of
eight cases depicted in fig. (4) {\it only two} yield a solution to Eq.
(\ref{3doch}) that can produce a finite $\sef$ value.  Therefore, these are
the only cases relevant for our discussion of bound states and resonances.
These cases are described by figs. (4b.ii) and (4d.i), which have the same
behaviour around $\si_2$ and $\si_3$.   As will be shown below, the finite
$\sef$ value for these cases is obtained in the limit
$\si_2\rta\si_1+\rta0$.  Two of the other cases in fig. (4) have already
been ruled out by Eq. (\ref{3dnev4}d) and the remaining four give rise to
infinite $\sef$'s or contain $\sigx$ configurations in euclidean time that
hit the $\si=0$ boundary point.

In both cases of interest to us here the quartic interaction is unbounded
from below ($\beta=-1$),  but as was discussed at the beginning of this
section, the field theory is perfectly defined to leading order in $1/N$ in
terms of the fluctuations around the vacuum of the theory which is a local
minimum of
$\sef$.  Though our resulting $\sigx$ configurations are quite distinct
from the vacuum condensation itself, they are of a finite amplitude, and
are therefore protected against being driven to infinity by the bottomless
quartic interaction.  The solution of Eq. (\ref{3doch}) for both these
cases read
\beq
\si(x) = \si_2+(\si_3-\si_2)cn^2\left[\sqrt{g(\si_3-\si_1)}\,
(x-x_0)\bmid\mu\right]  \label{3tcin}
\eeq
where $cn$ is a Jacobi elliptic function with parameter \cite{ww,ast}
\beq
\mu=\frac{\si_3-\si_2}{\si_3-\si_1}\;.  \label{3tsiz}
\eeq

For generic values of $\si_1,\si_2$ and $\si_3$ we have $0<\mu<1$ and
$\sigx$ has a finite real period $L$ along the $x$ axis given by
\beq
L=2K(\mu)  \label{3tset}
\eeq
where $K(\mu)$ is a complete elliptic integral of the first kind.  In
this case Eq. (\ref{3tcin}) yields obviously an infinite $\sef$ value,
unless the field theory is defined in a ``box'' of finite volume.  However,
for $\mu\rta  1-$,
$L$ in Eq. (\ref{3tset}) diverges as $\haf\,\ln\, \frac{1}{1-\mu}$ and
$\sigx$ degenerates into
\beq
\si(x)\mathop\approx\limits_{\mu\rta 1-} \si_2+\si_{32} \mbox{sech}^2\,
\left[\sqrt{g\si_{31}}\,(x-x_0)\right] + \cO(1-\mu) \label{3toch}
\eeq
where $\si_{3k}=\si_3-\si_k;\;k=1,2$.  Eq. (\ref{3toch}) leads evidently to a
finite $\sef$ value.

The desired limit $\mu\rta1-$ is obtained as $\si_1\rta\si_2-$.  In the
Minkowsky space theory (fig. (4b.ii)) this limit is possible only as both
$\si_1$ and $\si_2$  vanish, since they have opposite signs.  This is
possible only when
$\left(\frac{g}{2T}\right)^2$ in Eq. (\ref{3doch}) becomes very small.
In such a case we have from Eq. (\ref{3doch})
\beq
\left\{ \begin{array}{l}  \si_3 = {\displaystyle\frac{m^2}{g} + \cO\left(
\frac{g}{T^2}\right) \approx \frac{m^2}{g}} \\[8pt]
{\displaystyle{\si_2=-\si_1=\cO\left(\frac{\sqrt{g}}{T}\right) \approx 0}\;.}
\end{array} \right.  \label{3tnev}
\eeq
Using Eq. (\ref{3tnev}), Eq. (\ref{3toch}) becomes (for $\mu=1$)
\beq
g\si(x)=m^2 \mbox{sech}^2 [m(x-x_0)]\;.    \label{3curt}
\eeq

Strictly speaking the euclidean space theory approaches this limit through
complex $\si_2=\si_1^*$ values, and fig. (4d.i) does not hold in this case, but
the end result, Eqs. (\ref{3tnev}) and (\ref{3curt}) are the same, as is
obvious from Eq. (\ref{3doch}).

Substituting Eq. (\ref{3curt}) into Eq. (\ref{3bdos}) (for $\beta=-1,~
\al= \pm 1$) we find that $\bph$ is coupled to the one dimensional
potential
$V(x)$ given by
\beq
V(x) = m^2-2g\sigx = m^2-2m^2 \mbox{sech}^2[m(x-x_0)]\,.  \label{3cuna}
\eeq
$V(x)$ is a reflectionless Schr\"odinger potential with a {\it single}
bound state $\psi_0(x)$ at zero energy $\la_0=0$ \cite{refl}, where
\beq
\psi_0(x) = \sqrt{\frac{m}{2}}\; \mbox{sech}\; [m(x-x_0)]\;.  \label{3cdos}
\eeq
In addition to $\psi_0(x),\quad V(x)$ has a continuum of scattering states
\beq
\psi_q(x) = \left\{ iq-m \;\tanh\;\left[m(x-x_0)\right]\right\}e^{iqx}
\label{3ctre}
\eeq
with eigenvalues $\la_q^2=m^2+q^2$ in which the reflectionless nature of
$V(x)$ is explicit.

In this respect, our $\sigx$ configuration is reminiscent of the
Callan-Coleman-Gross-Zee kink in the Gross-Neveu model \cite{dhn3}.  As in
that kink solution, the profile of $\sigx$ is independent of the number
of particles ($\bph$ ``mesons'') trapped in it.  This is contrary to the
behaviour of $\sigx$ configurations found in \cite{dhn3,abbott} by inverse
scattering techniques, that are all small distortions of the vacuum
configuration $\si=\si_c$. We would like to stress at this point that unlike
the Callan-Coleman-Gross-Zee kink, our configuration (Eq. (\ref{3cuna})) does
not connect degenerate vacua, since the $\ON$ model lacks such structures.

As far as ``mesonic" {\it bound} states are concerned, we have only one
relevant quantum number $n_0$ in Eqs. (\ref{3bsiz})-(\ref{3bnev}),
corresponding to Eq. (\ref{3cdos}).  In this case the degeneracy factor in
Eq. (\ref{3bset}) reads
\beq
C(n_0) =\frac{(N+n_0-1)!}{(N-1)!\;n_0!} \label{3cqrt}
\eeq
which implies that the dimension of the multiplet with eigenvalue $E(n_0)$
is that of a symmetric tensor of rank $n_0$ which is a highly reducible
representation of $\ON$, the irreducible components of which are all
traceless symmetric tensors of lower ranks $n'_0$ such that
$n_0-n'_0\eqv0\,({\rm mod}\;2)$.

Up to this point we have not specified the value of $m$ in any of our
equations.  Explicit comparison with Eq. (4.38) of \cite{abbott} as
well as the analogy with the Callan-Coleman-Gross-Zee kink indicate that we
must identify $m$ as the finite (renormalized) mass of the $\bph$ ``mesons"
obtained from the ``gap-equation'' \cite{sh}.  Moreover, from Eq.
(4.14) of
\cite{abbott} it seems that our ``kink-like'' $\sigx$ configuration
corresponds to the parameter $\tha$ in that equation having the value
$\tha=\frac{\pi}{2}$.  Ref. \cite{abbott} considered only the range $
0\leq\tha\leq\frac{\pi}{4}$, discarding $\tha=\frac{\pi}{2}$.  The reason
for this was that only in that region did $E(n_0,\tha)$ obtained extrema
which were actually local minima of
$E(n_0,\tha)$ as a function of $\tha$.  But since we are discussing
excited states of the field theory, it is clear that these minima prevail
{\it only} in a subspace of the total Hilbert space of states that is
orthogonal to all lower energy states (recall Eqs.
(\ref{3tuna})-(\ref{3tqrt})).  Therefore there is nothing wrong with our
$\sigx$ configurations lying outside the domain considered by
\cite{abbott}.  It has been shown in \cite{abbott} that in the case of
stable quartic interactions $(\beta=+1)$ no saddle points $\sigx$
configurations that are minor distortions of the vacuum configuration
$\si=\si_c$ could be found.  We complement this conclusion by proving that
in this case ``big'' kink like $\sigx$ configurations do not occur as well.

We close this section by actually calculating the spectrum given by Eq.
(\ref{3bnev}) and (\ref{3tnta}).  The regularized lowest energy state in
the bound is given by
\beq
E_0^{\rm Reg}[\si]=E_0[\si]-E_0[m]\;,  \label{3ccin}
\eeq
Namely, following Eq. (\ref{3tnta}) (for $\beta=-1$)
\beq
\frac{1}{N}\,E_0^{\rm Reg}[\si]=\int\limits^\infty_{-\infty}\,\si^2\,dx
+\haf \sum_i\,\left(\la_i[\si]-\la_i[m^2]\right)\;. \label{3csiz}
\eeq
Note that $V(x)$ in Eq. (\ref{3cuna}) is isospectral to the ``potential''
$\til V(x)=m^2$, except for the {\it single} zero energy bound state of
$V(x)$
\cite{refl}.  Therefore, all terms in the sum over $\la_i$ in Eq.
(\ref{3csiz}) cancel identically except for $\la_0$, which is zero anyway.
Thus, only the integral in Eq. (\ref{3csiz}) contributes to $E_0^{\rm
Reg}$, yielding
\beq
E_0^{\rm Reg}\,[\si] = \frac{4Nm^3}{3g^2}\;.    \label{3cset}
\eeq

Since the only discrete eigenvalue of $V(x)$ in Eq. (\ref{3cuna}) is zero,
Eq. (\ref{3cset}) yields the common value of all bound state energies
\beq
E(n_0) =\frac{4Nm^3}{3g^2}  \label{3coch}
\eeq
independently of $n_0$.  In this case the whole band degenerates into a
single energy level.

\pagebreak
\baselineskip=20.5pt

\section*{Conclusion}

\vspace{-3pt}
In this work we have shown how simple properties of a one dimensional
Schr\"odinger hamiltonian can be used to obtain time or space dependent
saddle point configurations contributing to the path integral in the large
$N$ limit.  As a specific model we have analyzed the $\ON$ vector model in
$0+1$ and in $1+1$ space-time dimensions.

In the quantum mechanical case our results are equivalent {\it completely}
to those obtained directly from the radial Schr\"odinger equation in the
large
$N$ limit.  The time dependent $\si(t)$ configurations we have found are
nothing but the semiclassical radial trajectories of that equation.  In
this respect our calculation is to the radial Schr\"odinger equation
exactly the same as what the instanton calculus is to JWKB calculations of
tunnelling effects.

In the two dimensional field theoretic case our method has been applied to
find static saddle point configurations.  In this case, the static case was
approached through an ``adiabatic'' approximation that discarded all time
dependence.  This led to static saddle points that were quite distinct from
the homogeneous vacuum one and from other static configurations obtained as
minor distortions of the vacuum by use of inverse scattering techniques.
Many of our novel static saddle points are important only in finite volume
cases.  We did not elaborate on these in this paper.  Only one homogeneous
saddle point configuration turned out to be relevant in the case of
infinite volume.  The latter gave rise to ``mesonic" bound states at zero
binding energy and is reminiscent of the Callan-Coleman-Gross-Zee kink
found in the Gross-Neveu model.

All extremal $\sigx$ configuration we have found in the $\ON$ vector model
turn out to be saddle points of $\sef$ rather than local minima of that
object.  This, however should not be of any surprise, since our saddle
points correspond to excited states of the quantal system.

Our method could be applied straightforwardly to other two dimensional
field theoretic models.

\vspace{-3pt}
\section*{Acknowledgements}

\vspace{-3pt}
I would like to thank C.M. Bender, A. Duncan, W. Fischler, M.S. Marinov, M.
Moshe and S. Weinberg for valuable discussions and e-mail correspondence.
I am deeply indebted to Profs. Bender, Marinov and Moshe for their patience
and for the many things I have learned by discussing various aspects of
this work with them.  I would like also to thank C. Bender and the theory
group at the Washington University in St. Louis for their very kind
hospitality where part of this work has been done.  I would also like to
thank M. Shifman and A. Smilga for a discussion on a related problem.  This
research is supported by a Rothchild post-doctoral fellowship and also in
part by the Robert A. Welch foundation and NSF Grant Phy 9009850.

\pagebreak

\section*{Appendix }
\setcounter{equation}{0}

\renewcommand{\theequation}{A.\arabic{equation}}

{\bf The differential equation obeyed by the diagonal resolvent of a
Schr\"odinger operator in one dimension} \\

Consider the spectral problem for the one dimensional Schr\"odinger operator
\beq
\hat h=-\pa_x^2+U(x)
\eeq
on the segment $a\leq x\leq b$ (where any of the end points may be at infinity)
with boundary conditions
\beq
\al\psi (z)+\beta\psi'(z)=0;~~z=a,b
\eeq
on the wave functions.

The eigenfunctions $\psi_n(x)$ and the eigenvalues $E_n$ of $\hat h$ are
therefore the solutions of
\beq
\left[-\pa_x^2+U (x)\right]\,\psi(x)=E\,\psi(x)
\eeq
{\it subjected to the boundary condition of Eq. (A.2)}.  Let $E$ be any real
number, and let $\psi_a(x)\;(\psi_b(x))$ be the solution to Eq. (A.3) that
satisfy the boundary condition Eq. (A.2) only at $x=a\;(x=b)$, but not at the
other boundary point.

The Wronskian of $\psi_a$ and $\psi_b$
\beq
W_{ab} = \psi_a(x)\psi'_b(x)-\psi'_a(x)\psi_b(x) \eqv C(E)
\eeq
is $x$ independent, and clearly vanishes (as a function of $E$) {\bf only} on
the spectrum \break of $\hat h$ \footnote{Indeed, if $a$ and $b$ are finite,
the
spectrum of $\hat h$ is non-degenerate.  If either $a$ and/or $b$ become
infinite, the continuum part of the spectrum (if exists) is only two-fold
degenerate.  Explicitly, if $E$ is an eigenvalue, $\psi_a$ must satisfy the
boundary condition at $x=b$ as well, leading to $C(E)\eqv 0$.  Conversely, if
$C(E)=0,~\psi_a(x)$ satisfies the boundary condition at $x=b$, hence $E$ is in
the spectrum.}.  For a generic $E$, the Green function corresponding to Eq.
(A.3) is the symmetric function.
\beq
G_E(x,y) =\frac{1}{C(E)}\;\left[\Tha(x-y)\psi_a(y)\psi_b(x)+\Tha(y-x)
\psi_a(x)\psi_b(y)\right]
\eeq
where $\Tha(x)$ is the step function.

It is easy to check that $G_E(x,y)$ satisfies its defining equation
\beq
\left[-\pa_x^2+U(x)\right]\;G_E(x,y)=\left[-\pa_y^2 +U(y)\right]
G_E(x,y)=\del(x-y),
\eeq
and therefore
\beq
G_E(x,y)=\lag x|\frac{1}{-\pa^2 + \,U-E}\, |y\rag\;.
\eeq
The {\it diagonal} resolvent of $\hat h$ at energy $E$
\beq
R_E(x) = \lag x|\,\frac{1}{-\pa^2+U-E}\, |x\rag
\eeq
is obtained from $G_E$ as
\beq
R_E(x)=\lim_{\eps\rta 0+}\haf \left[ G_E(x,x+\eps)+ G_E(x+\eps,x)\right] =
\frac{\psi_a(x)\psi_b(x)}{C(E)}\,.
\eeq
Using Eq. (A.3) we find
\beq
R^{\prime\prime}_E(x)=2\left[(U-E) + \frac{\psi'_a}{\psi_a} \,
\frac{\psi'_b}{\psi_b}\right] \, R
\eeq
leading to
\beq
R^{\prime 2}_E - 2R_ER^{\prime\prime}_E=-4R_E^2(U-E)+
\left(\frac{W_{ab}}{\psi_a\psi_b}\right)^2 R^2\;.
\eeq
Using eqs. (A.4) and (A.9) we finally arrive at
\beq
-2R_E\,R^{\prime\prime}_E +R^{\prime 2}_E +4R_E^2(U-E)=1
\eeq
which is nothing but the ``Gelfand-Dikii'' equation \cite{gd}, alluded to in
sections II and III.

\pagebreak

\end{document}